\documentclass[epj,final]{svjour}
\usepackage{graphicx}
\usepackage{color}
\usepackage{amsmath,amsfonts,amssymb}
\numberwithin{equation}{section}
\begin{document}
\title{Two-dimensional perturbations  in a scalar model for shear-banding}
\author{Johan L.A. Dubbeldam\inst{1} \and P.D. Olmsted\inst{2}
}                     
\offprints{Johan L.A. Dubbeldam\\ \email{j.l.a.dubbeldam@ewi.tudelft.nl}}         
\institute{ Delft University of Technology,
  Mekelweg 4, 2628 CD Delft , The Netherlands\and Polymer IRC and School of Physics \& Astronomy, University of
Leeds, Leeds LS2 9JT, United Kingdom}
\date{Received:\today / Revised version: date}
%
\abstract{
We present an analytical study of a toy model for shear banding,
  without normal stresses, which uses a piecewise linear
  approximation to the flow curve (shear stress as a function of shear
  rate). This model exhibits multiple stationary states,  one of which is  
  linearly stable against general two-dimensional perturbations.  This
  is in contrast to analogous results for the Johnson-Segalman model,
  which includes normal stresses, and which has been reported to be
  linearly unstable for general two-dimensional perturbations. This
  strongly suggests that the linear instabilities found in the
  Johnson-Segalman can be attributed to normal stress effects.
} 
\maketitle

\begin{onecolumn}

\section{Introduction}
\label{sec:intro}

Shear-banding phenomena have received a lot of attention during the
last decade. In shear banding the material splits into different
spatial regions, or bands, that flow at different shear rates
$\dot{\gamma}$. This phenomenon has been observed in granular media
\cite{cohen2006sya} and in viscoelastic living polymer systems such as
wormlike micelle solutions \cite{Schm+94}.  Theoretically, shear
banding is fairly well understood at the level of a stable one
dimensional (1D) banding profile \cite{radulescu99b,SYC96,LOB00},
which connects two shear rates that are stable at a given selected
total shear stress $\Sigma$.  Wormlike micelles, polymer melts, and
liquid crystals naturally give rise to such bistable constitutive
relations \cite{doiedwards,SDO88,Cate96}.  However, the stability of
bands in two or three dimensions (i.e. with respect to capillary-like
fluctuations of the interface) is not well understood, despite a
frequent interpretation in terms of a simple stable flat interface
between coexisting states. It has been experimentally observed that
the interface between the two phases is not necessarily flat, but can
exhibit strong undulations and erratic fluctuations
\cite{Lopez-Gonzalez.Holmes.ea04,LerDecBer00,Becu.Manneville.ea04}.
Theoretical calculations have shown that in shear banding flows,
chaotic motion of shear bands can occur
\cite{CatHeaAjd02,Aradian.Cates05,fielding04}. Such chaotic motion has
also been inferred experimentally in recent works by Sood and
co-workers \cite{Ganapathy.Sood06,Ganapathy.Sood06*1,chaos2000}.

A recent numerical calculation in two dimensions, which was first performed by Fielding \cite{SMF2005} 
and whose results were later confirmed in Refs.~\cite{SMF2005,fielding06,WilsonFielding2006}, demonstrated that,
for the Johnson-Segalman model the interface between the low and high
shear rate phases is {\it linearly unstable} to undulations. In this
case the unstable mode involved normal stresses; however, it is still
not known whether normal stresses are \textit{carried by} an instability
inherent in the two-dimensional (2D) nature of the fluctuation, or
whether normal stresses \textit{trigger} the instability
\cite{WilsonFielding2006}.  Hence, in this paper we consider a simple
general toy model without normal stresses \cite{SYC96,fielding04}.  We
show that for a wide class of multivalued flow curves, a linear
instability can never occur. Our findings imply that the coupling
between convective terms and perturbations in shear stress cannot lead
to a linear instability, which suggests that an instability requires
other degrees of freedom, such as normal stresses.

This paper is organised as follows. In Section \ref{sec:model} we
introduce the model and study stationary solutions for general
constitutive equations in which the shear rate is multivalued for a
range of stresses.  2D perturbations are introduced in the model in
Section \ref{sec:stability} and a linear stability analysis is carried
out, from which we conclude that linear instability does not occur for
our system. This strongly suggests that normal stresses are
responsible for the linear instability observed in
Refs.~\cite{SMF2005}.  The results are discussed and summarized in
Section \ref{sec:discuss}.
\section{Model Description}
\label{sec:model}
We consider planar Couette shear flow between flat plates separated by
a distance $h$, for which the velocity field is $\vec{v} =
\dot{\gamma} y \hat{\vec{x}}$. In the very low Reynolds number limit,
which applies to the complex fluids of interest, the total shear
stress $\Sigma$ is uniform in space. We assume that ${\Sigma}$
comprises two terms,
\begin{equation}
  \label{eq:1}
{\Sigma}={\sigma}_p(y)+\eta{\dot{\gamma}}(y),
\end{equation}
where the second term is the Newtonian stress of the solvent and
${\sigma}_p$ is the polymer stress.  In this paper, we only consider
the shear component of the stress tensor, but take the 2D nature of
the perturbations in the flow field into account. Following
\cite{fielding04,radulescu99b,SYC96}, we consider the following
governing equation for the polymer shear stress:
\begin{equation}
\left(\partial_t+{\bf
v}\cdot\nabla\right)\sigma_p\!=\!-\frac{\sigma_p}{\tau}+\frac{G}{\tau}
g({\dot{\gamma}}\tau)+D\nabla^2\sigma_p,
\label{eq:2}
\end{equation}
where $\tau$ is the relaxation time of the polymer stress, $D$
is the stress diffusion coefficient, and $G$ is the plateau modulus.  The
function $g$ is nonmonotonic, and in Refs.~\cite{SYC96,fielding04} was
taken to be
\begin{equation}
   g(\xi)=\frac{\xi}{1+\xi^2}.
\label{eq:g}
\end{equation}
Stress diffusion is necessary to define a uniquely selected stress
$\Sigma^{\ast}$~\cite{LOB00}, as is usually found in experiments under
controlled shear rate conditions.  Moreover, it is natural that
spatial gradients in the microstructure should be penalized, no matter
how weakly.

We next make all quantities dimensionless, as noted by a carat,
by expressing stress relative to $G$, length relative to the
plate separation $h$, and time relative to the relaxation time $\tau$:
\begin{align}
 \hat{\sigma}_p&=\sigma/G& \widehat{\Sigma}&=\Sigma/G \\
\hat{\dot{\gamma}}&=\tau\dot{\gamma}&
 \hat{D}&=D\tau/h^2\\
\hat{y}&=y/h & \hat{x}&=x/h\\
\hat{t}&=t/\tau &
\hat{v}&=v\tau/h
\end{align}


The total stress is a nonmonotonic
function of shear rate $\hat{\dot{\gamma}}$.  For an average shear
rate $\langle{\hat{\dot{\gamma}}}\rangle$ imposed on the unstable
portion of the flow curve, with negative slope
$\partial\widehat{\Sigma}/\partial\hat{\dot{\gamma}}<0$, the system
will break up into two bands, and hence a spatial variation along
$\hat{y}$. Upon integrating Eq.~(\ref{eq:1}) over the vertical
distance between the plates we find
\begin{equation}
  \label{eq:Sigma}
\widehat{\Sigma}=\langle \hat{\sigma}_p\rangle+\frac{1}{\alpha}\langle
{\hat{\dot{\gamma}}}\rangle,
\end{equation}
where spatial averages are defined by
\begin{equation}
 \langle {\cal{ O}} \rangle=\int_0^1{\cal O}(\hat{y})\,d\hat{y}
\end{equation}
and we have introduced the dimensionless parameter
where $\alpha\equiv G\tau/\eta$.
For an inhomogeneous steady state, the polymer stress equation,
Eq.~(\ref{eq:2}), leads to
\begin{equation}
-\hat{\sigma}_p+g(\alpha(\widehat{\Sigma}-\hat{\sigma}_p))+\hat{D}
\hat{\sigma}_p''=0,
\label{eq:stat1}
\end{equation}
where the strain rate has been eliminated using Eq.~(\ref{eq:1}) and
the prime denotes a $\hat{y}$-derivative,
$\hat{\sigma}_p'=\partial{\hat{\sigma}_p}/{\partial\hat{y}}$. As in
\cite{SMF2005,olmsted99a}, we impose Neumann boundary conditions,
$\hat{\sigma}_p'(\hat{y}=0) = \hat{\sigma}_p'(\hat{y}=1) = 0$, for
which  Eq.~(\ref{eq:stat1})  can be solved numerically for a given
stress $\widehat{\Sigma}$.

The selected stress $\widehat{\Sigma}^{\ast}$ is determined by the
constraint of \textit{fixed average shear rate} $\langle
\hat{\dot{\gamma}}\rangle$.  We thus eliminate $\widehat{\Sigma}$ from
the stationary condition, Eq.~(\ref{eq:stat1}), using
Eq.~(\ref{eq:Sigma}) leading to
\begin{equation}
-\hat{\sigma}_p+g\left[\alpha \left(\langle\hat{\sigma}_p\rangle
 -\hat{\sigma}_p\right) +
 \langle\hat{\dot{\gamma}}\rangle\right]+\hat{D}\hat{\sigma}_p''=0.
\label{eq:stat2}
\end{equation}
We numerically solve this integro-differential equation
using a semi-implicit Crank-Nicolson algorithm~\cite{numC}. Once a
solution is found, the selected stress $\widehat{\Sigma}^{\ast}$ is obtained
using Eq.~(\ref{eq:Sigma}). Fig.~\ref{fig:fig1} shows  banding
profiles $\hat{\sigma}_p(\hat{y})$ for the function $g(\xi) $ given by
Eq.~(\ref{eq:g}).

\begin{figure}[t]
\includegraphics[width=8.40truecm]{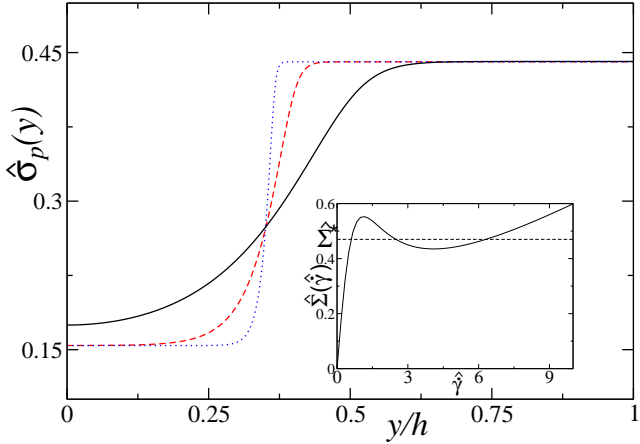}
\caption{The polymer stress $\sigma_p$ as a function of the distance
  $\hat{y}=y/h$ from the lower plate for $\hat{D}=0.0001$ (dotted),
  $\hat{D}=0.001$ (dashed) and $\hat{D}=0.01$ (solid), for $\alpha=20$
  and average shear rate $\langle\tilde{{\gamma}}\rangle=2.60$.  The
  selected stress is $\widehat{\Sigma}^{\ast}=0.470$ for
  $\hat{D}=0.001$ and $\hat{D}=0.0001$; and
  $\widehat{\Sigma}^{\ast}=0.471$ for $\hat{D}=0.01$. The inset shows
  the corresponding constitutive curve, shear stress
  $\widehat{\Sigma}$ as a function of $\hat{\dot{\gamma}}$, for
  $\alpha=20$.}
\label{fig:fig1}
\end{figure}

\begin{figure}
\includegraphics[width=8.0truecm]{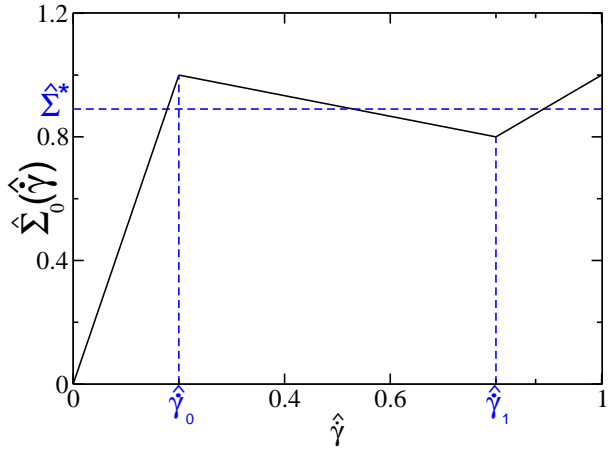}
\caption{ The selected stress $\widehat{\Sigma}^{\ast}$ and the piecewise
  linear function $\widehat{\Sigma}_0(\hat{\dot{\gamma}})$ as
  a function of $\hat{\dot{\gamma}}$ for $\hat{\dot{\gamma}}_0=0.2$,
  $\hat{\dot{\gamma}}_1=0.8$, $A_c=5$; the selected stress is
  $\widehat{\Sigma}^{\ast}=0.891$.}
\label{fig:fig2}
\end{figure}

\begin{figure}
\includegraphics[width=7.5cm]{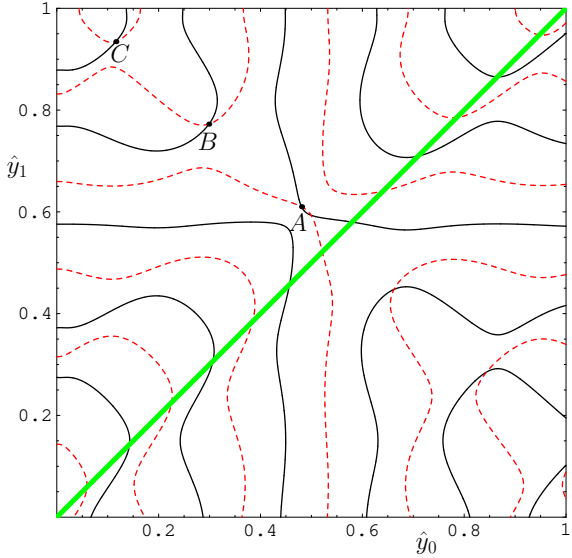}
\caption{The solutions $A$, $B$, $C$, for $\hat{y}_0$ and $\hat{y}_1$,
  for the parameter values: $A_c=5, \alpha=1,
  \langle{\hat{\dot{\gamma}}}\rangle=0.50, \hat{\dot{\gamma}}_0=0.2,
  \hat{\dot{\gamma}}_1=0.8$, and $\hat{D}=0.001$. The dashed curves
  correspond to Eq.~(\ref{eq:c3}) and the solid ones to
  Eq.~(\ref{eq:c4}). The physical range of solutions is in the upper
  left triangle, $0< y_0< y_1< 1$.}
\label{fig:fig3}
\end{figure}


\begin{figure}[t]
\includegraphics[width=6.80cm]{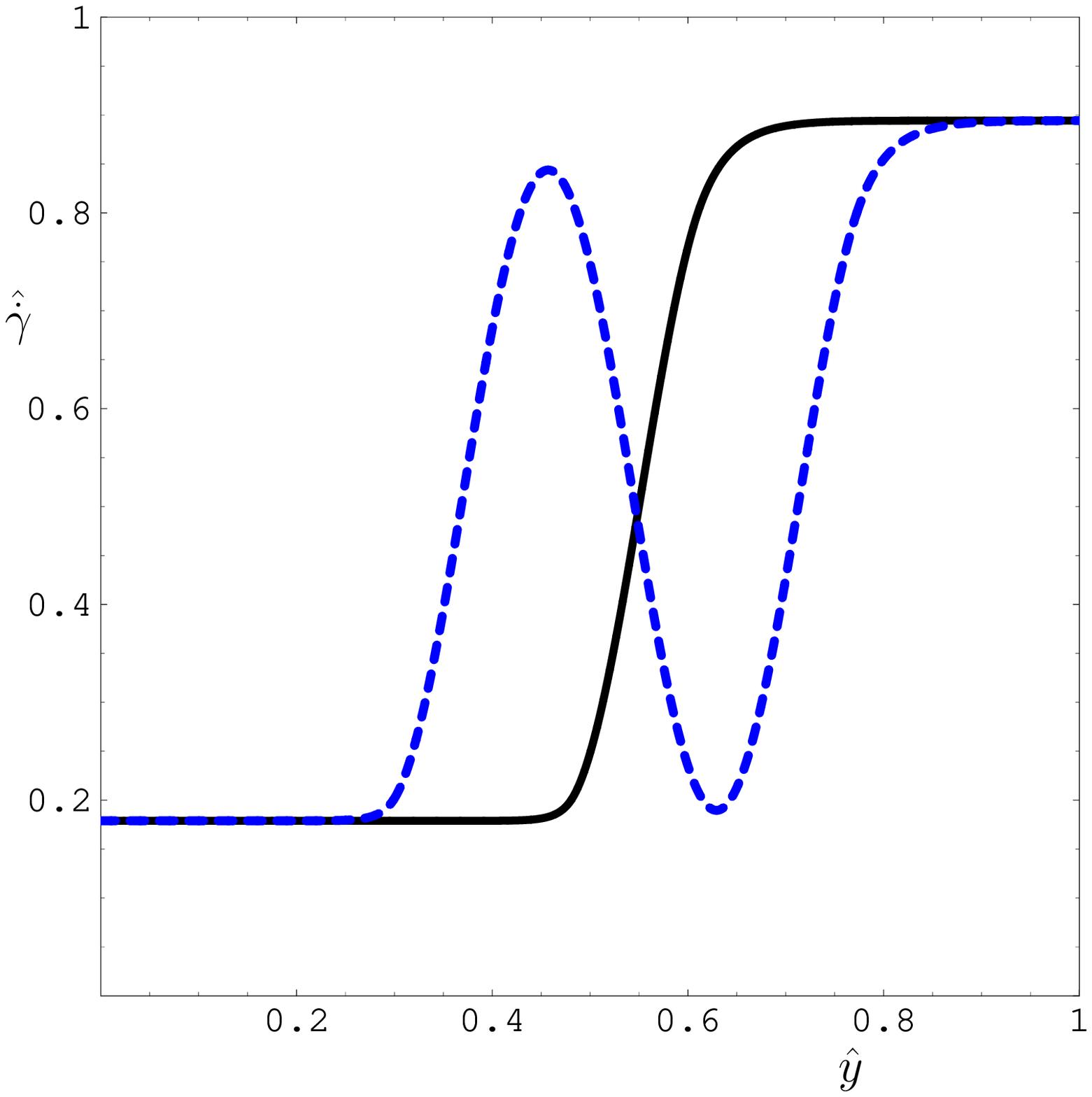}
\caption{Two stationary solutions corresponding to $A$ (solid curve)
 and $B$ (dashed curve). An additional third solution with one more
 oscillation is not shown.}
\label{fig:fig4}
\end{figure}

To better understand stress selection we will generalize $g(\xi)$ to a
function that allows explicit analytic calculations, but is
sufficiently general to describe shear banding. The calculations
discussed in the main text concern this general model.  However, for
comparison we also reproduce the calculations for the function $g$
defined in Eq.~(\ref{eq:g}), in Appendix \ref{sec:contmodel0}.  The
most essential characteristic of $g$ is that it vanishes for small and
large shear rates, so that the function $\widehat{\Sigma}(x) =
x/\alpha+g(x)$ is nonmonotonic with a maximum and a minimum.  A simple
piecewise linear function showing this behavior is
\begin{eqnarray}
\widehat{\Sigma}_0(\hat{\dot{\gamma}})=
\begin{cases}
\displaystyle  \frac{A_c \hat{\dot{\gamma}}}{\alpha} &
  (\hat{\dot{\gamma}}<\hat{\dot{\gamma}}_0)\\[10truept]
\displaystyle  \left[\frac{A_c\hat{\dot{\gamma}}_0 - \hat{\dot{\gamma}}_1}
  {\hat{\dot{\gamma}}_0
      - \hat{\dot{\gamma}}_1}\right] \frac{(\hat{\dot{\gamma}} -
    \hat{\dot{\gamma}}_0)}{\alpha} + 
  \frac{A_c\hat{\dot{\gamma}}_0}{\alpha} &
  (\hat{\dot{\gamma}}_0<\hat{\dot{\gamma}}<\hat{\dot{\gamma}}_1)\\[10truept]
\displaystyle\frac{\hat{\dot{\gamma}}}{\alpha} &
(\hat{\dot{\gamma}}>\hat{\dot{\gamma}}_1), 
\end{cases}
\label{eq:H}
\end{eqnarray}
where $A_c\hat{\dot{\gamma}}_0 > \hat{\dot{\gamma}}_1$.  The form of
$\widehat{\Sigma}_0(\hat{\dot{\gamma}})$ is completely specified by
the three parameters $\hat{\dot{\gamma}}_0$, $\hat{\dot{\gamma}}_1$, and
$A_c$.

Using $\widehat{\Sigma}_0(\hat{\dot{\gamma}}) -
\hat{\dot{\gamma}}/\alpha$ instead of $g$ in the equation of motion
for $\hat{\sigma}_p$, and eliminating $\hat{\sigma}_p$ using
Eq.~(\ref{eq:1}), the steady state condition for
$\hat{\dot{\gamma}}$ is given by
\begin{equation}
 \hat{D}\hat{\dot{\gamma}}''+\alpha\widehat{\Sigma}=
 \alpha\widehat{\Sigma}_0(\hat{\dot{\gamma}}).
\end{equation}
This is solved with Neumann boundary conditions for the stress at
$\hat{y}=0$ and $\hat{y}=1$, leading to
\begin{subequations}
\begin{align}
\hat{\dot{\gamma}}_I(\hat{y}) &=
c_1\cosh\left(\sqrt{\frac{A_c}{\hat{D}}}\hat{y}\right) +
\alpha\frac{\widehat{\Sigma}}{A_c}&& (0<\hat{y}<\hat{y}_0)\label{eq:w1}\\
\hat{\dot{\gamma}}_{II}(\hat{y})&=d_1\cos\left(\sqrt{\frac{z}{\hat{D}}}
  {\hat{y}}\right) +
d_2\sin\left(\sqrt{\frac{z}{\hat{D}}}{\hat{y}}\right)+\hat{\dot{\gamma}}_0 - \frac{(A_c\hat{\dot{\gamma}}_0 -
  \alpha\widehat{\Sigma})}{A_c\hat{\dot{\gamma}}_0 -
  \hat{\dot{\gamma}}_1}(\hat{\dot{\gamma}}_0 - \hat{\dot{\gamma}}_1)
&&(\hat{y}_0<\hat{y}<\hat{y}_1)
\label{eq:w2}\\
\hat{\dot{\gamma}}_{III}(\hat{y})&=\alpha\widehat{\Sigma} +
c_2\cosh\left(\frac{\hat{y}-1}{\sqrt{\hat{D}}}\right)
&&(\hat{y}_1<\hat{y}<1)\label{eq:w3}
\end{align}
\end{subequations}
where
$z\equiv(A_c\hat{\dot{\gamma}}_0-
\hat{\dot{\gamma}}_1)/(\hat{\dot{\gamma}}_1-\hat{\dot{\gamma}}_0)$, which can be recognized 
as the negative of the slope of the function $\widehat{\Sigma}_0(\hat{\dot{\gamma}})$ between $\hat{\dot{\gamma}}_0$ and $\hat{\dot{\gamma}}_1$.
The profile obeys the following continuity conditions:
\begin{align}
\hat{\dot{\gamma}}_{I}(\hat{y}_0)&=\hat{\dot{\gamma}}_{II}(\hat{y}_0)=
\hat{\dot{\gamma}}_0\label{eq:c1}\\
\hat{\dot{\gamma}}_{II}(\hat{y}_1)&=\hat{\dot{\gamma}}_{III}(\hat{y}_1)=
\hat{\dot{\gamma}}_1\label{eq:c2}\\
\hat{\dot{\gamma}}_{I}'(\hat{y}_0)&=\hat{\dot{\gamma}}'_{II}(\hat{y}_0)\label{eq:c3}\\
\hat{\dot{\gamma}}_{II}'(\hat{y}_1)&=\hat{\dot{\gamma}}'_{III}(\hat{y}_1),\label{eq:c4}
\end{align}
where again the prime denotes a $y$-derivative,
$\hat{\dot{\gamma}}'=\partial \hat{\dot{\gamma}}/\partial \hat{y}$.
These six conditions guarantee a smooth solution for the shear rate
and stress and from them the four constants $c_1$, $c_2$, $d_1$, $d_2$
and the locations $\hat{y}_0$ and $\hat{y}_1$ where the function
$\widehat{\Sigma}_0(\hat{\dot{\gamma}})$ switches from one piece to
the next, can be determined.  Appendix \ref{sec:intconst} contains
expressions for the constants $c_1$, $c_2$, and $d_1$, $d_2$ as functions of $\hat{y}_0$ and $\hat{y}_1$.
$d_1$ and $d_2$ are found by combining Eqs.~(\ref{eq:c1}) and (\ref{eq:c3}).
 Eqs.~(\ref{eq:c2}) and (\ref{eq:c4}) are then used to determine the values of $\hat{y}_0$ and $\hat{y}_1$. The
selected stress $\widehat{\Sigma}^{\ast}$ is determined by the constraint of
an applied average shear rate $\langle{\hat{\dot{\gamma}}}\rangle$,
which is  given by Eq.~(\ref{eq:ap3}).
In a shear banding state the selected stress $\widehat{\Sigma}^{\ast}$
satisfies
$\hat{\dot{\gamma}}_1<\widehat{\Sigma}^{\ast}<A_c\hat{\dot{\gamma}}_0$, and
we must have $\hat{\dot{\gamma}}>0$ on the interval $\hat{y}\in[0,1]$.


A graphical solution for $\hat{y}_0$ and $\hat{y}_1$, for the case
$\hat{D}=0.001$, $A_c=5$ and $\langle\hat{\dot{\gamma}}\rangle=0.5$,
is depicted in Fig.~\ref{fig:fig3}, as the intersections of the
contour plots of Eqs.~(\ref{eq:c3}) (dashed line) and (\ref{eq:c4})
(solid line), which should simultaneously be obeyed by $\hat{y}_0$ and
$\hat{y}_1$. To solve these equations the expressions for the
constants $c_1,c_2,d_1,d_2$ from Appendix \ref{sec:intconst} are used,
Eqs.~(\ref{eq:constants}), as well as expression (\ref{eq:ap3}) for
the selected stress.  The physical range of Fig.~\ref{fig:fig3} is the
upper left hand triangle $0<\hat{y}_0<\hat{y}_1<1$, to the left of
the diagonal line in Fig.~\ref{fig:fig3}.

Fig.~\ref{fig:fig3} shows {\it three} intersection points
corresponding to stationary solutions, denoted by $A=(0.4806,0.6101)$,
$B=(0.2984,0.7726)$, and $C=(0.1160,0.9345)$. The values of the
selected stress $\widehat{\Sigma}^{\ast}$ associated with these
solutions are $0.8944$ for $A$, $0.8946$ for $B$ and $0.8949$ for $C$.
The fourth intersection point (the companion of $A$, with the
unphysical value of $\widehat{\Sigma}^{\ast}=4.8$) is ignored, as it
violates the condition $\hat{\dot{\gamma}}_1<\widehat{\Sigma}<A_c
\hat{\dot{\gamma}}_0$.



The first solution $A$ corresponds to the common shear rate profile
depicted as a solid curve in Fig.~\ref{fig:fig4}.  The stationary
solution $B$ (shown dashed) has an extra oscillation in the interface
region (Fig.~\ref{fig:fig4}), while the third solution $C$ has two
oscillations (not shown). The selected stress changes monotonically as
a function of an increasing number of oscillations in the resulting
solution.  The existence of multiple solutions for small diffusion
constants is well-known in the context of pattern formation in
reaction-diffusion models \cite{grindrod}.  Multiple solutions occur
in our model when $\hat{D}\lesssim10^{-2}$, with the multiplicity
increasing for decreasing $\hat{D}$.  Having obtained the stationary
solutions we next examine their stability by performing a linear
perturbation analysis.

\section{Linear stability analysis}
\label{sec:stability}
\subsection{Governing Equations and Matching Conditions}
\label{sec:governing-equations}



We study the linear stability of the piecewise linear model by
introducing small perturbations to the stationary 1D solutions,
\begin{align}
\hat{\sigma}_p(\hat{x},\hat{y})&=\hat{\sigma}_p^{1D}(\hat{y}) +
\delta\hat{\sigma}_p(\hat{x},\hat{y})\\
\label{eq:18}
\widehat{\Sigma}(\hat{x})&=\widehat{\Sigma}^{1D} +
\delta\widehat{\Sigma}(\hat{x})\\
\hat{\dot{\gamma}}(\hat{x},\hat{y})&=\hat{\dot{\gamma}}^{1D}(\hat{y})
+ \delta\hat{\dot{\gamma}}(\hat{x},\hat{y})\\
\hat{\boldsymbol{v}}&=(\hat{v}^{1D}(\hat{y}) + \delta
\hat{v}_{\hat{x}}(\hat{x},\hat{y}),\delta \hat{v}_{\hat{y}}(\hat{x},\hat{y})),
\label{eq:vfield}
\end{align}
where the superscript $^{1D}$ denotes the stationary 1D solution.
Perturbations in the shear rate, polymer stress and total stress are
related by
$\delta\widehat{\Sigma}=\delta\hat{\sigma}_p+\hat{\dot{\gamma}}/\alpha$,
which can be recast as 
\begin{equation}
\langle\delta\hat{\sigma}_p\rangle=\delta\hat{\sigma}_p+\delta\hat{\dot{\gamma}}/\alpha
\label{eq:5}
\end{equation}
by averaging and using $\langle\delta\hat{\dot{\gamma}}\rangle=0$. Note that $\langle\delta\hat{\dot{\gamma}}\rangle=0$
implies that the average perturbation in the polymer stress $\langle\delta\hat{\sigma}_p\rangle$ equals the 
perturbation in the total stress  $\delta\widehat{\Sigma}$, for $\hat{\dot{\gamma}}=\alpha(\delta\widehat{\Sigma}-\delta\hat{\sigma}_p)$. 
 The  shear rate
is given by
$\hat{\dot{\gamma}}(x,y)=\partial_{{\hat{y}}}(\hat{v}^{1D}(\hat{y})+\delta
\hat{v}_x(\hat{x},\hat{y}))$, so that
$\delta\hat{\dot{\gamma}}(\hat{x},\hat{y})=\partial_{\hat{y}}\delta
v_{\hat{x}}(\hat{x},\hat{y})$.  Because the flow is 
incompressible, a variation of the velocity in the $x$-direction must
be accompanied by a compensating variation of the velocity in the
$y$-direction, which obeys
\begin{equation}
\partial_{\hat{y}}\delta \hat{v}_y(\hat{x},\hat{y})=-\partial_{\hat{x}}\delta \hat{v}_x(\hat{x},\hat{y}).
\label{eq:incomp}
\end{equation}

We recall that the equation of motion of the polymer stress is given by
\begin{equation}
(\partial_{\hat{t}}+\hat{\bf{v}}\cdot\nabla)\hat{\sigma}_p=\widehat{\Sigma}_0(\hat{\dot{\gamma}})-\widehat{\Sigma}+\hat{D}\nabla^2\hat{\sigma}_p.
\end{equation}
This equation can be recast in a differential equation for the shear rate $\hat{\dot{\gamma}}$,
by using $\delta\hat{\sigma}_p=\delta\widehat{\Sigma}-\delta\hat{\dot{\gamma}}/\alpha$, from which we find to first 
order in all perturbations 

\begin{multline}
\label{eq:pert}
\partial_{\hat{t}}\delta\hat{\dot{\gamma}}+\hat{v}^{1D}(\hat{y})
\partial_{{\hat{x}}}\delta\hat{\dot{\gamma}}+\delta
\hat{v}_{\hat{y}}\partial_{{\hat{y}}}\hat{\dot{\gamma}}^{1D}(\hat{y}) = (\partial_{\hat{t}}+\hat{v}_{\hat{x}}^{1D}\partial_{\hat{x}})\alpha\delta\widehat{\Sigma}\\
 + \hat{D}\left[\partial_{\hat{x}\hat{x}}+\partial_{\hat{y}\hat{y}}\right]\delta\hat{\dot{\gamma}}
-\hat{D}\partial_{\hat{x}\hat{x}}\alpha\delta\widehat{\Sigma} 
-\alpha\widehat{\Sigma}_0(\hat{\dot{\gamma}}^{1D}+\delta\hat{\dot{\gamma}})+
\alpha\widehat{\Sigma}_0(\hat{\dot{\gamma}}^{1D}),
\end{multline}
where we explicitly kept $\widehat{\Sigma}_0(\hat{\dot{\gamma}}^{1D}+\delta\hat{\dot{\gamma}})$ and 
$\widehat{\Sigma}_0(\hat{\dot{\gamma}}^{1D})$  in order to avoid differentiating the function
$\widehat{\Sigma}_0$ with respect to $\hat{\dot{\gamma}}$. This was done as the function $\widehat{\Sigma}_0$ is  continuous
for all $\hat{\dot{\gamma}}$, but non-differentiable  in the turning points of the stress at $\hat{\dot{\gamma}}_0,
\hat{\dot{\gamma}}_1$.  

 Using Eqs.~(\ref{eq:5}) and (\ref{eq:incomp}) we can easily express the
 perturbed velocities $\delta \hat{v}_x(\hat{x},\hat{y})$ and $\delta
 \hat{v}_y(\hat{x},\hat{y})$ in terms of $\delta\hat{\dot{\gamma}}$.
 Starting from the identity $\frac{\partial\delta
   v_{\hat{x}}(\hat{x},\hat{y},\hat{t})}{\partial
   \hat{y}}=\delta\hat{\dot{\gamma}}$, we integrate from $0$ to
 $\hat{y}$ and use the no slip boundary condition at $\hat{y}=0$, which yields

\begin{align}
\delta \hat{v}_x(\hat{x},\hat{y})=
\int_0^{\hat{y}}\delta\hat{\dot{\gamma}}(\hat{x},y')dy'.
\label{eq:dvx}
\end{align}
If one next differentiates $ \delta \hat{v}_x(\hat{x},\hat{y})$ with
respect to $\hat{x}$ and substitutes the resulting expression in
Eq.~(\ref{eq:incomp}), only one integration with respect to $\hat{y}$
(again starting from $0$ and using no slip boundary conditions)
is necessary to obtain the perturbed velocity in the $\hat{y}$ direction
\begin{align}
\delta \hat{v}_y(\hat{x},\hat{y}) =& 
  -\int_0^{\hat{y}}\left[\int_0^{y'}\partial_{\hat{x}}\delta\hat{\dot{\gamma}}
  (\hat{x},y'')dy''\right]\,dy'.\label{eq:dvy} 
\end{align}
Hence the equation for the
shear rate perturbation $\delta\hat{\dot{\gamma}}$ (\ref{eq:pert}) can
be expressed entirely in terms of $\delta\hat{\dot{\gamma}}$, integrals
over $\delta\hat{\dot{\gamma}}$ and $\delta\widehat{\Sigma}$.
Perturbations in the shear rate must be
continuous across the matching points of the bands,
which are yet to be found. These follow from demanding
\begin{align}
\label{eq:deltag0}
\left.\left(\widehat{\dot{\gamma}}^{1D} +
    \delta\widehat{\dot{\gamma}}\right)
\right|_{\hat{y}_0+\delta{\hat{y}_0}}=\widehat{\dot{\gamma}}_0,\,\,\,\,\,\,\,  
\left.\left(\widehat{\dot{\gamma}}^{1D}+\delta\widehat{\dot{\gamma}}
  \right)\right|_{\hat{y}_1+\delta{\hat{y}_1}}=\widehat{\dot{\gamma}}_1, 
\end{align}
that is, the matching points are shifted by perturbations $\delta
\hat{y}$ needed to bring the local shear rate to the matching points
in the constitutive relation $\Sigma_0(\hat{\dot{\gamma}})$.  Hence we
split the interval $[0,1]$ into three parts: $(I)$,
$y\in[0,\tilde{y}_0]$; $(II)$, $y\in[\tilde{y}_0,\tilde{y}_1]$, and
$(III)$, $y\in[\tilde{y}_1,1]$, where
$\tilde{y}_i=\hat{y}_i+\delta\hat{y}_i(\hat{x},\hat{t}), i=1,2$.  The
equations of motion for the perturbations in the polymer stress are,
from Eq.~\eqref{eq:pert}
\begin{subequations}\label{eqs:alldse}
\begin{align}
\label{seq:dtsigma1}
\partial_{\hat{t}}\delta\hat{\dot{\gamma}}^{I}&=\hat{D}\nabla^2\delta\hat{\dot{\gamma}}^{I}
-\hat{v}_{\hat{x}}^{1D}(\hat{y})\partial_{\hat{x}}\delta\hat{\dot{\gamma}}^{I}-\delta
\hat{v}_{\hat{y}}(\hat{x},\hat{y})\partial_{\hat{y}}\hat{\dot{\gamma}}^{1D}
 - A_c\delta\hat{\dot{\gamma}}^{I}+F(\alpha\delta\widehat{\Sigma},\hat{t})\\[8truept]
\label{seq:stsigma2}
\partial_{\hat{t}}\delta\hat{\dot{\gamma}}^{II}&=\hat{D}\nabla^2\delta\hat{\dot{\gamma}}^{II}
-\hat{v}_{\hat{x}}^{1D}(\hat{y})\partial_{\hat{x}}\delta\hat{\dot{\gamma}}^{II}-\delta
\hat{v}_{\hat{y}}(\hat{x},\hat{y})\partial_{\hat{y}}\hat{\dot{\gamma}}^{1D}
+z\delta\hat{\dot{\gamma}}^{II}+F(\alpha\delta\widehat{\Sigma},\hat{t})\\[8truept] 
\label{seq:dtsigma3}\partial_{\hat{t}}\delta\hat{\dot{\gamma}}^{III}&=\hat{D}\nabla^2\delta\hat{\dot{\gamma}}^{III}
-\hat{v}_{\hat{x}}^{1D}(\hat{y})\partial_{\hat{x}}\delta\hat{\dot{\gamma}}^{III}-\delta\hat{v}_{\hat{y}}(\hat{x},\hat{y})\partial_{\hat{y}}\hat{\dot{\gamma}}^{1D}
-\delta\hat{\dot{\gamma}}^{III}+F(\alpha\delta\widehat{\Sigma},\hat{t})
\end{align}
\end{subequations}
where we defined $F(\alpha\delta\widehat{\Sigma},\hat{t})=(\partial_{\hat{t}}+\hat{v}_{\hat{x}}^{1D}\partial_{\hat{x}}-\hat{D}\partial_{\hat{x}\hat{x}})\alpha\delta\widehat{\Sigma}(\hat{x},\hat{t})$. 
Eqs.~(\ref{eqs:alldse})  must be supplemented with boundary conditions the matching points. At $\hat{y}=\tilde{y}_0$ this
 condition can be found from Eq.~(\ref{eq:deltag0}),
\begin{align}
 \hat{\dot{\gamma}}^{1D}(\hat{y}_0+\delta\hat{y}_0(\hat{x},\hat{t})) +
 \delta\hat{\dot{\gamma}}(\hat{y}_0+\delta\hat{y}_0(\hat{x},\hat{t}))=
 \hat{\dot{\gamma}}_0,
\end{align}
which leads, to first order in the perturbed quantities, to 
\begin{align}
\label{eq:deltay0}
 \delta\hat{\dot{\gamma}}(\hat{y}_0,\hat{x},\hat{t}) =
 -\left(\frac{\partial\hat{\dot{\gamma}}^{1D}}{\partial\hat{y}}
 \right)_{\hat{y}_0}\delta\hat{y}_0(\hat{x},\hat{t}).  
\end{align}
Eq.~(\ref{eq:deltay0}) relates the change in the shear rate to a shift in the 
interface position.  
The boundary condition at $\hat{y}=\tilde{y}_1$ is derived similarly. From Eq.~(\ref{eq:deltay0}) we 
immediately retrieve the shift in the value of $y_0$, and hence the new position of the two shear bands, once $\delta\hat{\dot{\gamma}}$ is known.
 To deal with the inhomogeneous (time dependent) boundary conditions in region $(II)$ we
define a new function $q(\hat{x},\hat{y},\hat{t})$ by
\begin{align}
\label{eq:defq}
q(\hat{x},\hat{y},\hat{t})&=\delta\hat{\dot{\gamma}}^{II}
(\hat{x},\hat{y},\hat{t})+\frac{(\hat{y}-\hat{y}_1)\delta\hat{\dot{\gamma}}(\hat{y}_0,\hat{x},\hat{t})}{\hat{y}_1-\hat{y}_0}-\frac{(\hat{y}-\hat{y}_0)\delta\hat{\dot{\gamma}}(\hat{y}_1,\hat{x},\hat{t})}{\hat{y}_1 -\hat{y}_0}\,{\equiv}\, \delta\hat{\dot{\gamma}}^{II}(\hat{x},\hat{y},\hat{t})+G(\hat{x},\hat{y},\hat{t}),
\end{align}
which satisfies continuity at the matching points.  Hence, the
function $q(\hat{x},\hat{y},\hat{t})$ vanishes at the boundary points
by construction. In Eq.~(\ref{eq:defq}) we introduced the function $G(\hat{x},\hat{y},\hat{t})$ 
for notational convenience. Using  Eq.~\eqref{seq:stsigma2}, we find the following
differential equation for $q$:
\begin{align}
\label{eq:qt}
&\partial_{\hat{t}} q =z q +\hat{D}\nabla^2 q -z  G+F(\alpha\delta\widehat{\Sigma},\hat{t})+F(G,\hat{t})-\hat{v}_{\hat{x}}^{1D}(\hat{y})\partial_{\hat{x}}q- \delta \hat{v}_{\hat{y}}(\hat{x},\hat{y})\partial_{\hat{y}}
\hat{\dot{\gamma}}^{1D}.
\end{align}

Now that we have derived the governing equations for the perturbations, we study their 
stability properties by employing a Fourier expansion in spatial coordinates.
\subsection{Fourier Expansion}
\label{sec:fourier-expansion}
We next expand the shear rate perturbation $\delta\hat{\dot{\gamma}}$
in Fourier modes, within the three regions, consistent with the
boundary conditions $\hat{\dot{\gamma}}'=0$ at $\hat{y}=0$ and
$\hat{y}=1$. This results in a \emph{continuous but non-smooth} function
$\delta\hat{\dot{\gamma}}$ \emph{over the entire interval} $[0,1]$:
\begin{subequations}\label{eq:allds}
\begin{align}
\delta\hat{\dot{\gamma}}^{I}(\hat{x},\hat{y},\hat{t})&=
\sum_{n=0}^{\infty}\cos\left(\frac{\pi
    \hat{y}(2n+1)}{2\tilde{y}_0}\right)A_n(\hat{x},\hat{t}) +
\delta\hat{\dot{\gamma}}(\hat{y}_0,\hat{x},\hat{t})
&& (0<\hat{y}<\tilde{y}_0)\label{seq:ds1}\\ 
q(\hat{x},\hat{y},\hat{t})&=\sum_{n=1}^{\infty}B_n(\hat{x},\hat{t})
\sin\left(\frac{\pi n(\hat{y}-\hat{y}_0)}{\hat{y}_1 -
    \hat{y}_0}\right)
&& (\tilde{y}_0<\hat{y}<\tilde{y}_1)\label{eq:6}\\
\delta\hat{\dot{\gamma}}^{III}(\hat{x},\hat{y},\hat{t})&=
\sum_{n=0}^{\infty}\cos\left(\frac{\pi
    (\hat{y}-1)(2n+1)}{2(1-\tilde{y}_1)}\right)D_n(\hat{x},\hat{t}) +
\delta\hat{\dot{\gamma}}(\hat{y}_1,\hat{x},\hat{t})
&&(\tilde{y}_1<\hat{y}<1).\label{seq:ds3} 
\end{align}
 \end{subequations}
 Only odd cosine modes contribute in regions $I$ and $III$ because we
 require zero derivatives at the ends of the interval $[0,1]$ and a
 matching condition at the right, respectively left, ends of the
 intervals $[0,\tilde{y}_0]$ and $[\tilde{y}_1,1]$. If one substitutes
 $\hat{y}=\hat{y}_0$ in Eq.~(\ref{seq:ds1}) one finds
 $\delta\hat{\dot{\gamma}}^{I}=\delta\hat{\dot{\gamma}}(\hat{y}_0,\hat{x},\hat{t})$, which coincides with the boundary value of
 $\delta\hat{\dot{\gamma}}$ in $\hat{y}_0$, derived above in
 Eq.~(\ref{eq:deltag0}), which guarantees continuity of $\delta\hat{\dot{\gamma}}(\hat{x},\hat{y},\hat{t})$ throughout the interval $[0,1]$

One can easily demonstrate that the solution of Eqs.~(\ref{eq:allds}) is non-smooth, by trying to match the derivatives at $\hat{y}=\hat{y}_0,\,\hat{y}_1$ using
Eqs.~(\ref{eq:allds}). We will not reproduce these calculation here, but only note that the narrower the interface, that is the smaller $\hat{y}_1-\hat{y}_0$, the closer the matching at the the interface points $\hat{y}_0$ and $\hat{y}_1$. The piecewise linear model will in this limit approach the smooth model defined by Eq.~(\ref{eq:g}). 

We therefore conclude that the perturbations are in general only continuous. To 
obtain an equation linking $\delta\hat{\dot{\gamma}}(\hat{y}_0,\hat{x},\hat{t})$ and $\delta\hat{\dot{\gamma}}(\hat{y}_1,\hat{x},\hat{t})$ we invoke the  constraint $\langle\hat\delta{\dot{\gamma}}\rangle=0$,
from which we find
\begin{align}
\label{eq:dg0dg1}
\frac{\delta\hat{\dot{\gamma}}(\hat{y}_0,\hat{x},\hat{t})}{2}\left(\hat{y}_0+\hat{y}_1\right)+\frac{\delta\hat{\dot{\gamma}}(\hat{y}_1,\hat{x},\hat{t})}{2}\left(2-\hat{y}_1-\hat{y}_0\right)=-\sum_{n=0}^{\infty}\frac{A_n(-1)^n 2y_0}{(2n+1)\pi}-\sum_{n=0}^{\infty}\frac{2(1-\hat{y}_1)D_n(-1)^n}{(2n+1)\pi}-\!(\hat{y_1}-\hat{y}_0)\sum_{n=1}^{\infty}\frac{B_n(1-(-1)^n)}{n\pi}.
\end{align}
Eq.~(\ref{eq:dg0dg1}) constitutes the  final equation needed to close the system, in a way which is similar to the
method we used in section \ref{sec:model} to deduce the value of $\widehat{\Sigma}$, as we will now demonstrate.

The general idea is to find the governing equation for the amplitudes $A_n$, $B_n$, $D_n$ from Eqs.~(\ref{eqs:alldse}) by projecting out the
cosine (for $A_n$ and $D_n$) and sine modes (for $B_n$). The evolution for  $\delta\hat{\dot{\gamma}}(\hat{y}_1,\hat{x},\hat{t})$ and $\delta\hat{\dot{\gamma}}(\hat{y}_0,\hat{x},\hat{t})$ can be obtained by integrating Eqs.~(\ref{seq:dtsigma1}) and (\ref{seq:dtsigma3})
over their respective intervals. This is most easily done in the Fourier domain, so 
we expand $A_n(\hat{x},\hat{t})$, $B_n(\hat{x},\hat{t})$,
$D_n(\hat{x},\hat{t})$, $\delta\widehat{\Sigma}(\hat{x},\hat{t})$, and  $\delta\hat{\dot{\gamma}}(\hat{y}_0,\hat{x},\hat{t})$ and $\delta\hat{\dot{\gamma}}(\hat{y}_1,\hat{x},\hat{t})$ in Fourier transforms: 
\begin{align}
A_n(\hat{x},\hat{t})&=\int_{-\infty}^{\infty}a_n(k,\hat{t})e^{ik\hat{x}}\,dk,&
B_n(\hat{x},\hat{t})&=\int_{-\infty}^{\infty}b_n(k,\hat{t})e^{ik\hat{x}}\,dk,&
D_n(\hat{x},\hat{t})&=\int_{-\infty}^{\infty}d_n(k,\hat{t})e^{ik\hat{x}}\,dk,&\nonumber\\
\delta\widehat{\Sigma}(\hat{x},\hat{t})&=\int_{-\infty}^{\infty}\delta\widehat{\Sigma}(k,\hat{t})e^{ik\hat{x}}\,dk,&
\delta\hat{\dot{\gamma}}(\hat{y}_0,\hat{x},\hat{t})&=\int_{-\infty}^{\infty}\delta\hat{\dot{\gamma}}(\hat{y}_0,\hat{k},\hat{t})e^{ik\hat{x}}\,dk,&\delta\hat{\dot{\gamma}}(\hat{y}_1,\hat{x},\hat{t})&=\int_{-\infty}^{\infty}\delta\hat{\dot{\gamma}}(\hat{y}_1,\hat{k},\hat{t})e^{ik\hat{x}}\,dk.
\label{eq:Aan}
\end{align}
From Eqs.~(\ref{seq:dtsigma1}) and (\ref{seq:ds1}) we then obtain the
following evolution equation for the amplitudes $a_n(k,\hat{t})$ and $\delta\hat{\dot{\gamma}}(\hat{y}_0,k,\hat{t})$
The amplitudes $a_n(k,\hat{t})$ can be extracted from this equation by
projecting out the different cosine modes, whereas direct integration gives 
the evolution equation for $\delta\hat{\dot{\gamma}}(\hat{y}_0,k,\hat{t})$.
We can perform the same calculations on $\delta\hat{\dot{\gamma}}^{III}$,
and $q$ which leads to similar expressions involving $d_n(k,\hat{t})$, $\delta\hat{\dot{\gamma}}(\hat{y}_1,k,\hat{t})$, 
and $b_n(k,\hat{t})$, which are derived in Appendix
\ref{sec:matderive}. 
It turns out that the introduction of scaled amplitudes ($\alpha_n$, $\beta_n$, $\delta_n$) is beneficial 
for numerical and notational reasons; they and related to ($a_n$, $b_n$, $d_n$)  according to
\begin{align}
\label{eq:scaled}
\alpha_n=\frac{a_n(-1)^n(2n+1)\pi}{4},&\hspace{12mm} \beta_n=\frac{b_n n \pi}{4},& \delta_n=\frac{d_n(-1)^n(2n+1)\pi}{4}.
\end{align}

To show the form of the evolution equation of the amplitudes, we  reproduce the  governing equation for the scaled amplitudes $\alpha_n(k,\hat{t})$
\begin{align}
\label{eq:govanR}
&\frac{d\alpha_m}{d\hat{t}}-\left(\frac{\partial\hat{\dot{\gamma}}^{1D}}{\partial\hat{y}}\right)_{\hat{y}_0}\partial_{\hat{t}}\delta \hat{y}_0(k,\hat{t})-\partial_{\hat{t}}\alpha\delta\widehat{\Sigma}(k,\hat{t})= (A_c+\hat{D}k^2)\left(\frac{\partial\hat{\dot{\gamma}}^{1D}}{\partial\hat{y}}\right)_{\hat{y}_0}\delta \hat{y}_0(k,\hat{t})-\left(A_c+\hat{D}k^2+\hat{D}\left(\frac{\pi(2m+1)}{2\hat{y}_0}\right)^2\right)\alpha_m\nonumber\\
&+\hat{D}k^2\alpha\widehat{\Sigma}(k,\hat{t})-\frac{ik(-1)^m(2m+1)\pi}{2\hat{y}_0}\left[\sum_{n=1}^{\infty}\alpha_n(-1)^n\frac{16\hat{y}_0^2}{(2n+1)^3\pi^3}\left[R^A_{mn}-T^A_m\right] +\left(\frac{\partial\hat{\dot{\gamma}}^{1D}}{\partial\hat{y}_0}\right)_{\hat{y}_0}\delta\hat{y}_0(k,\hat{t}) F^A_m\right.\nonumber\\
&+\left.\sum_{n=1}^{\infty}\left[L^A_{mn}+\frac{D}{A_c}R^A_{mn}\right]\frac{4\alpha_n(-1)^n}{(2n+1)\pi} -\left(J^A_m+\frac{\hat{D}}{A_c}T^A_m\right)\left[\alpha\delta\widehat{\Sigma}+\left(\frac{\partial\hat{\dot{\gamma}}^{1D}}{\partial\hat{y}}\right)_{\hat{y}_0}\delta\hat{y}_0(k,\hat{t})\right]\right].\nonumber\\
\end{align}
The quantities $F^A_m$, $R^A_{mn}$, $T^A_m$, $L^A_{mn}$, $J^A_m$ are defined in terms of integrals over the interval $[0,\hat{y}_0]$. Their precise definitions can be found in Appendix \ref{sec:matderive}, Eqs.~(\ref{eq:integralcoef}).

Finally, we need one further differential equation for the evolution of $\delta\widehat{\Sigma}(\hat{x},\hat{t})$.
This equation is found by differentiating the identity (\ref{eq:dg0dg1}) with respect to time and subsequently substituting the expressions for $\partial_{\hat{t}}\delta\hat{\dot{\gamma}}(\hat{y}_0,k,\hat{t})$ (Eq.~(\ref{eq:govg0})) and $\partial_{\hat{t}}\delta\hat{\dot{\gamma}}(\hat{y}_1,k,\hat{t})$
(Eq.~(\ref{eq:govg1}). This yields
\begin{align}
\label{eq:govdS}
&\alpha\partial_{\hat{t}}\delta\widehat{\Sigma}-4(\hat{y}_1-\hat{y}_0)\sum_{n=0}^{\infty}\frac{d\alpha_n}{dt}\frac{1}{(2n+1)^2\pi^2}-4(\hat{y}_1-\hat{y}_0) \sum_{n=0}^{\infty}\frac{d\delta_n}{dt}\frac{1}{(2n+1)^2\pi^2}+8(\hat{y}_1-\hat{y}_0)\sum_{n=0}^{\infty}\frac{d\beta_{2n+1}}{d\hat{t}}\frac{1}{(2n+1)^2\pi^2}=\nonumber\\
&+4(1+\hat{D}k^2)(2-\hat{y}_1-\hat{y}_0)\sum_{n=0}^{\infty}\frac{\delta_n}{(2n+1)^2\pi^2}+
4(A_c+\hat{D}k^2)(\hat{y}_1+\hat{y}_0)\sum_{n=0}^{\infty}\frac{\alpha_n}{(2n+1)^2\pi^2}+\frac{\hat{D}(\hat{y}_0+\hat{y}_1)}{\hat{y}_0^2}\sum_{n=0}^{\infty}\frac{\alpha_n}{(2n+1)^2\pi^2}\nonumber\\
&+\frac{\hat{D}(2-\hat{y}_0-\hat{y}_1)}{(1-\hat{y}_1)^2}\sum_{n=0}^{\infty}\frac{\delta_n}{(2n+1)^2\pi^2}+(1+\hat{D}k^2)(2-\hat{y}_1-\hat{y}_0)\frac{\delta\hat{\gamma}_1}{2}+(A_c+\hat{D}k^2)(\hat{y}_1+\hat{y}_0)\frac{\delta\hat{\gamma}_0}{2}-\alpha\hat{D}k^2\delta\widehat{\Sigma}+ik\Pi(k,\hat{t}),
\end{align}
where $\Pi(k,\hat{t})$ contains the imaginary contribution to (\ref{eq:govdS}), whose exact form can be found in Appendix \ref{sec:matderive}.

To summarize, the linear stability of the shear banding states to undulating perturbation is investigated 
by analyzing the evolution of the amplitudes $\alpha_n(k,\hat{t})$, $\beta_n(k,\hat{t})$, $\delta_n(k,\hat{t})$,  $\delta\widehat{\Sigma}(k,\hat{t})$ and $\delta\hat{\dot{\gamma}}(\hat{y}_0,\hat{x},\hat{t})$, $\delta\hat{\dot{\gamma}}(\hat{y}_1,\hat{x},\hat{t})$ governed by Eqs.~(\ref{eq:govanR}),~(\ref{eq:govbn}),~(\ref{eq:govdn}), 
(\ref{eq:govg0}), (\ref{eq:govg1}) and (\ref{eq:govdS}). The profile, and hence the interface,  is stable only if all these amplitudes  decay with time.

\subsection{Non-undulating perturbations for $\hat{D}{\to}0$}\label{subsec:d0}
Before turning to the general case, we will examine the limit of
vanishing $\hat{D}$ and $k=0$. This special case was investigated for
the Johnson-Segalman model in \cite{fielding04} and was also explored
in \cite{yih}. The conclusion in both cases was that the shear bands 
in the system are
neutrally stable in the absence of diffusion terms. 

 \begin{figure}[t]
   \includegraphics[width=7.80cm]{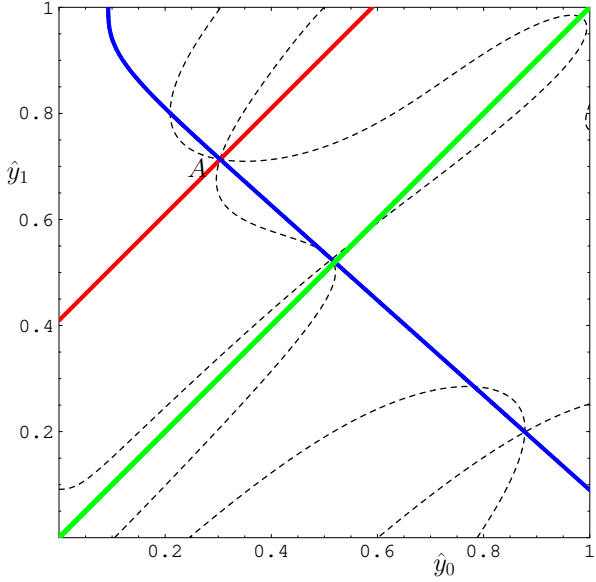}
   \caption{Comparison of the analytical solution for $\hat{y}_0$ and $\hat{y}_1$, depicted  
by the intersection of the two solid curves given by (Eq.~(\ref{eq:y0ana})) and (Eq.~(\ref{eq:Dsqa})), with the solution obtained by the method of section~\ref{sec:model},
which are printed dashed. The parameter values are $A_c=5.0$, $\hat{\dot{\gamma}}_0=0.2$, $\hat{\dot{\gamma}}_1=0.8$, $\langle\hat{\dot{\gamma}}\rangle=0.5$ and $\hat{D}=0.01$. The diagonal curve indicates the line $\hat{y}_1=\hat{y}_0$ and physical
solutions are in the upper left triangle. The intersection point $A$ between the two dashed curves
is practically on top of the intersection point of the solid curves ($\hat{y}_0=0.3033$,$\hat{y}_1=0.7130$).}
\label{fig:anasol}
 \end{figure}


 \begin{figure}[t]	
   \includegraphics[width=8.20cm]{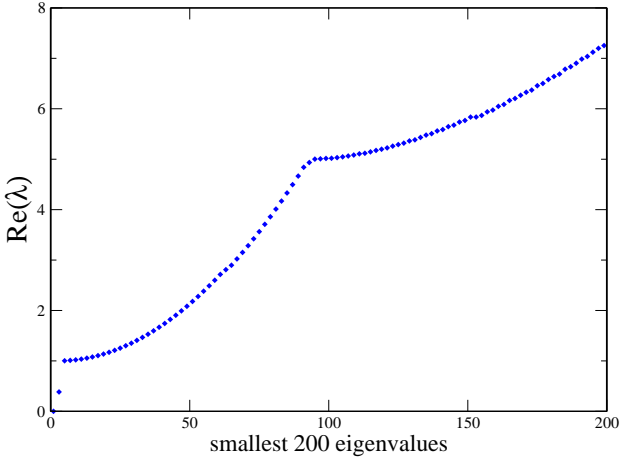}
   \caption{Real parts of the 200 eigenvalues with smallest real parts of the system of equations (\ref{eqs:allk0}) for non-undulating perturbations.
   The parameter values are  $A_c=5, \hat{y}_0=0.4806, \hat{y}_1=0.6101,
     \langle\hat{\dot{\gamma}}\rangle=0.5, \hat{\dot{\gamma}}_0=0.2,
     \hat{\dot{\gamma}}_1=0.8$, and $\hat{D}=10^{-5}$. The smallest eigenvalue is $0$, demonstrating that the system is 
neutrally stable. Eigenvalues are more concentrated around $\lambda=1$ and $\lambda=5$ in agreement with the analytical asymptotic result.The first eigenvalue $\lambda=0$ is related to $\delta\widehat{\Sigma}$, the second with $\lambda\,\approx\,0.1$ to $\beta_1$ and all eigenvalues $\lambda=1$ to $\alpha_n$, and $\lambda=A_c=5$ to $\delta_n$.}
\label{fig:fig5}
 \end{figure}


 In section \ref{sec:model} we  demonstrated how to find the
 stationary state(s) of system defined by Eq.~(\ref{eq:2}) graphically
 as the intersection point(s) of two families of curves,
 defined by Eqs.~(\ref{eq:c3}) and (\ref{eq:c4}).  For the case $\hat{D}\to 0$, we can
 actually find an excellent analytic approximation to the stationary solution(s) which is
remarkably close to stationary solutions found by the graphical method.  
 In the limit  $\hat{D}\to 0$ equations (\ref{eq:c2}) and (\ref{eq:c4}) reduce to
\begin{subequations}
\label{eq:Dz}
\begin{align}
\frac{\alpha\widehat{\Sigma}-A_c\hat{\dot{\gamma}}_0}{z}
\cos\left(\sqrt{\frac{z}{\hat{D}}}(\hat{y}_1-\hat{y}_0)\right) 
 + \left(\hat{\dot{\gamma}}_0 -
  \frac{\alpha\widehat{\Sigma}}{A_c}\right)\sqrt{\frac{A_c}{z}}\sin\left(\sqrt{\frac{z}{\hat{D}}}
  (\hat{y}_1-\hat{y}_0)\right)&=\left(1+\frac{\alpha\widehat{\Sigma}-A_c\hat{\dot{\gamma}}_0}{A_c
    \hat{\dot{\gamma}}_0-\hat{\dot{\gamma}}_1}\right)
(\hat{\dot{\gamma}}_1-\hat{\dot{\gamma}}_0)
\label{eq:Dz1}\\
\frac{\alpha\widehat{\Sigma}-A_c\hat{\dot{\gamma}}_0}{z}\sin\left(\sqrt{\frac{z}{\hat{D}}}
  (\hat{y}_1-\hat{y}_0)\right)
-\sqrt{\frac{A_c}{z}}\left(\hat{\dot{\gamma}}_0 -
  \frac{\alpha\widehat{\Sigma}}{A_c}\right)\cos\left(\sqrt{\frac{z}{\hat{D}}}(\hat{y}_1-\hat{y}_0)\right)
 &=\frac{(\hat{\dot{\gamma}}_1-{\alpha\widehat{\Sigma}})}{\sqrt{z}},
\label{eq:Dz2}
\end{align}
\end{subequations}
keeping all terms up to order $\sqrt{\hat{D}}$.  If we next square Eqs.~(\ref{eq:Dz1}) 
and (\ref{eq:Dz2}) and add them we find the following relation 
\begin{align}
 \label{eq:Dsqa}
\left(\frac{\alpha\widehat{\Sigma}-A_c\hat{\dot{\gamma}}_0}{\alpha\widehat{\Sigma}-\hat{\dot{\gamma}}_1}\right)^2&=\frac{A_c(z+1)}{z+A_c}.
\end{align}
If we would next substitute the value of $\widehat{\Sigma}$ as given by Eq.~(\ref{eq:ap3}), we could express $\hat{y}_1$ in terms
of $\hat{y}_0$. However, first we obtain the width of the interface, that is the value of $\hat{y}_1-\hat{y_0}$, to verify if it
scales proportionally to ${\sqrt{\hat{D}}}$ in agreement with findings in the literature. 
We therefore solve Eqs.~(\ref{eq:Dz}) for $\sin\left(\sqrt{\frac{z}{\hat{D}}}(\hat{y}_1-\hat{y}_0)\right)$ 
and obtain, after using Eq.~(\ref{eq:Dsqa}),
\begin{align}
\label{eq:sin}
\sin\left(\sqrt{\frac{z}{\hat{D}}}(\hat{y}_1-\hat{y}_0)\right)=\frac{\sqrt{A_c}+1}{\sqrt{z+1+A_c+A_c/z}}.
\end{align}

When we invert this equation we get the following expression for interface width

\begin{align}
\label{eq:y1my0}
\hat{y}_1-\hat{y}_0=\sqrt{\frac{\hat{D}}{z}}\left((2M+1)\pi-\arcsin\left(\frac{\sqrt{A_c}+1}{\sqrt{z+1+A_c+A_c/z}}\right)\right).
\end{align}
The other solution of Eq.~(\ref{eq:sin}) is not compatible with the condition $\hat{\dot{\gamma}}_1<\alpha\widehat{\Sigma}<A_c\hat{\dot{\gamma}}_0$ and therefore does 
not correspond to a physical solution. The index $M$ is the band index. $M=0$ designates the 
two band solution ($A$), $M=1$ has two extra bands, like solution $B$, and so on.  
Expression (\ref{eq:y1my0}) confirms that the interface scales as $\sqrt{\hat{D}}$, as
observed in \cite{olmsted99a,fielding04}, and gives an exact value for
the interface length as a function of $z$ and $A_c$.
An implication of Eq.~(\ref{eq:y1my0}) is that shear bands can
only exist when $\hat{D}$ is sufficiently small. Indeed the maximum
possible width of the interface is $\hat{y}_1-\hat{y}_0\simeq 1$, implying a
maximum value of $\hat{D}^{\ast}\,\approx\,\frac{z}{(\pi-1)^2}$. In the case we have been
considering $z=1/3$, and $A_c=5$, we find  $\hat{D}^{\ast}\simeq 0.054$,
which agrees with our numerical findings. 
The values of $\hat{y}_0$ and $\hat{y}_1$ for small $\hat{D}$ can be determined analytically
by using Eq.~(\ref{eq:Dsqa}) once more, and substituting the expression of $\alpha\widehat{\Sigma}^{\ast}$
as given in Eq.~(\ref{eq:ap3}) thereby using Eq.~(\ref{eq:y1my0}). This calculation  gives the following expression for
$\hat{y}_0$, which designates the position of the interface of the steady state with $2(M+1)$ shear bands 
\begin{align}
\label{eq:y0ana}
&\hat{y}_0=\left[\frac{(A_c-1)(\sqrt{A_c\hat{\dot{\gamma}}_0}+\sqrt{\hat{\dot{\gamma}}_1}}{\hat{\dot{\gamma}}_0\hat{\dot{\gamma}}_1}\right]\left((\hat{\dot{\gamma}}_1-\langle\hat{\dot{\gamma}}\rangle)\sqrt{A_c\hat{\dot{\gamma}}_0}+(A_c\hat{\dot{\gamma}}_0-\langle\hat{\dot{\gamma}}\rangle)\sqrt{\hat{\dot{\gamma}}_1}-\sqrt{{\hat{D}}}\left(1+\frac{(2M+1)\pi-h}{\sqrt{z}}\right)(A_c-1)\right),
\end{align}
where  $h\,\equiv\,\arcsin\left(\frac{\sqrt{A_c}+1}{\sqrt{z+1+A_c+A_c/z}}\right)\,\simeq\,0.77$ for the parameter values
used in this paper. From Fig.~\ref{fig:anasol}, one can see that the analytical solution (\ref{eq:y0ana})  with $M=0$ found for $\hat{D}\to0$, is 
still an excellent approximation for $\hat{D}=0.01$. The other stationary states with multiple ($2(M+1)$) bands can be found similarly, by 
substituting the corresponding value of $M$ in Eq.~(\ref{eq:y0ana}). Note that multiple bands are not present for $\hat{D}=0.01$, as is shown in Fig.~\ref{fig:anasol} where we only find one intersection point ($A$) of the solid curves. 

\begin{figure}
\label{fig:eigenval}
      \includegraphics[width=8.00cm]{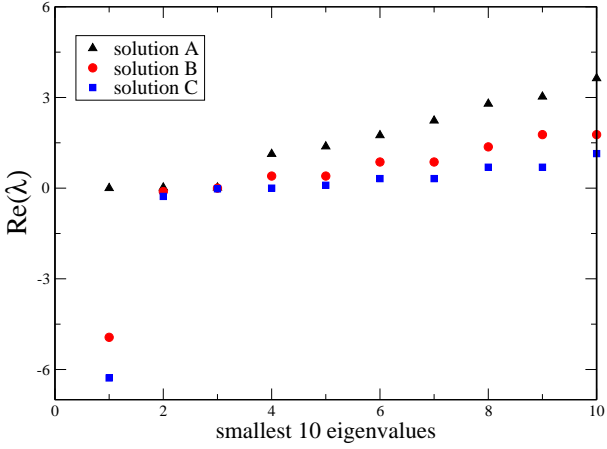}
      \caption{The real part of the first 10 eigenvalues of ${\bf S}^{-1}{\bf L^R}$ for an
        expansion of Eq.~(\ref{eq:1}) in $n=1000$ modes with zero wave
        number $k$ for solutions A $(\triangle)$, B $(\circ)$, and C
        $(\square)$. The parameters are the same as in Fig.~3:
        $A_c=5$, $\langle{\hat{\dot{\gamma}}}\rangle=0.50$, and
        $\hat{\dot{\gamma}}_0=0.2$, $\hat{\dot{\gamma}}_1=0.8$; the
        diffusion constant $\hat{D}=0.001$. It can be clearly seen
        that the real part of the eigenvalues for solutions B and C
        cross zero, indicating instability of the solutions with
        multiple bands. Only the solution with two bands (A) is (neutrally)
        stable.}
\label{fig:figST}
\end{figure}
\vskip 0.2cm

In order to study the stability of the solutions obeying
Eq.~(\ref{eq:y1my0}), to non-undulating perturbations  we need to determine the eigenvalues of the
evolution equations for $a_n(\hat{t})$, $b_n(\hat{t})$, $d_n(\hat{t})$ (or equivalently $\alpha_n(\hat{t})$, $\beta_n(\hat{t})$, $\delta_n(\hat{t})$), $\delta\hat{\dot{\gamma}}(\hat{y}_0,\hat{t})$, $\delta\hat{\dot{\gamma}}(\hat{y}_0,\hat{t})$ and $\delta\widehat{\Sigma}(\hat{t})$  for $k=0$. Because the imaginary parts of the governing equations all depend linearly on $k$,  we can
restrict ourselves to the real part of the differential equations for the amplitudes  $\alpha_n$, $\beta_n$, $\delta_n$,  $\delta\hat{\dot{\gamma}}(\hat{y}_0,\hat{t})$, $\delta\hat{\dot{\gamma}}(\hat{y}_0,\hat{t})$ and $\delta\widehat{\Sigma}$. These equations can easily be found by projecting out the different sine and cosine modes and are given by Eqs.~(\ref{eq:govan}), (\ref{eq:govbn}), (\ref{eq:govdn}), (\ref{eq:govg0}),  (\ref{eq:govg1}), and (\ref{eq:govsig}).
For $k=0$ and $\hat{D}\to0$ the evolution equations for $\alpha_n(k,\hat{t})$, $\beta_n(k,\hat{t})$, $\delta_n(k,\hat{t})$, $\delta\hat{\dot{\gamma}}(\hat{y}_0,\hat{t})$, $\delta\hat{\dot{\gamma}}(\hat{y}_1,\hat{t})$  and  $\delta\widehat{\Sigma}(k,\hat{t})$ read:
\begin{subequations}\label{eqs:allk0}
\begin{align}
&\frac{d\alpha_m}{d\hat{t}}-\frac{d\alpha\delta\widehat{\Sigma}}{d\hat{t}}+\frac{d\delta\hat{\dot{\gamma}}(\hat{y}_0,\hat{t})}{d\hat{t}}=-A_c \delta\hat{\dot{\gamma}}(\hat{y}_0,\hat{t})-A_c \alpha_m \label{eq:Ak0}\\
&\frac{d\beta_m}{d\hat{t}}-\frac{d\alpha\delta\widehat{\Sigma}}{d\hat{t}}\left(\frac{1-(-1)^m}{2}\right)+\frac{1}{2}\frac{d\delta\hat{\dot{\gamma}}(\hat{y}_0,\hat{t})}{d\hat{t}}-\frac{(-1)^m}{2}\frac{d\delta\hat{\dot{\gamma}}(\hat{y}_1,\hat{t})}{d\hat{t}}
=z\left(1-\frac{m^2\pi^2}{((2M+1)\pi-h)^2}\right)\beta_m\nonumber\\
&+\frac{z}{2}\delta\hat{\dot{\gamma}}(\hat{y}_0,\hat{t})-\frac{z(-1)^m}{2}\delta\hat{\dot{\gamma}}(\hat{y}_1,\hat{t})
\label{eq:Bk0}\\
&\frac{d\delta_m}{d\hat{t}}-\frac{d\alpha\delta\widehat{\Sigma}}{d\hat{t}}+\frac{d\delta\hat{\dot{\gamma}}(\hat{y}_1,\hat{t})}{d\hat{t}}=- \delta\hat{\dot{\gamma}}(\hat{y}_1,t)- \delta_m\label{eq:Dk0}\\
&\frac{d\delta\hat{\dot{\gamma}}(\hat{y}_0,\hat{t})}{d\hat{t}}-\frac{d\alpha\delta\widehat{\Sigma}}{d\hat{t}}+\sum_{n=0}^{\infty}\frac{d\alpha_n}{d\hat{t}}\frac{8}{(2n+1)^2\pi^2}=-A_c\delta\hat{\dot{\gamma}}(\hat{y}_0,\hat{t})-A_c\sum_{n=0}^{\infty}\frac{8\alpha_n}{(2n+1)^2\pi^2}\label{eq:g0k0}\\
&\frac{d\delta\hat{\dot{\gamma}}(\hat{y}_1,\hat{t})}{d\hat{t}}-\frac{d\alpha\delta\widehat{\Sigma}}{d\hat{t}}+\sum_{n=0}^{\infty}\frac{d\delta_n}{d\hat{t}}\frac{8}{(2n+1)^2\pi^2}=-\delta\hat{\dot{\gamma}}(\hat{y}_1,\hat{t})-\sum_{n=0}^{\infty}\frac{8\delta_n}{(2n+1)^2\pi^2}\label{eq:g1k0}\\
&\frac{d\alpha\delta\widehat{\Sigma}}{d\hat{t}}=A_c(\hat{y}_0+\hat{y}_1)\frac{\delta\hat{\dot{\gamma}}(\hat{y}_0,\hat{t})}{2}+(2-\hat{y}_1-\hat{y}_0)\frac{\delta\hat{\dot{\gamma}}(\hat{y}_1,\hat{t})}{2}+4A_c(\hat{y}_0+\hat{y}_1)\sum_{n=0}^{\infty}\frac{\alpha_n}{(2n+1)^2\pi^2}+4(2-\hat{y}_1-\hat{y}_0)\sum_{n=0}^{\infty}\frac{\delta_n}{(2n+1)^2\pi^2}\label{eq:Sk0}.
\end{align}
\end{subequations}
Eqs.~(\ref{eqs:allk0}) contain a lot of information. It is immediately clear, for example, that $\lambda=0$ is an
eigenvalue of Eqs.~(\ref{eqs:allk0}), by which we mean that solutions  with all amplitudes decaying as $\exp(-\lambda t)$ with $\lambda=0$, exist. This can be seen by interpreting Eqs.~(\ref{eqs:allk0}) as a matrix equation which has
the eigenvalue corresponding to $\delta\widehat{\Sigma}$ equal to zero, 

If it is proved that all amplitudes except $\frac{d\delta\widehat{\Sigma}}{d\hat{t}}=0$ are always decay, and thus have eigenvalues with positive real parts , we 
have shown that the shear banded state is neutrally stable. This agrees
with Fielding \cite{fielding04} and Yih \cite{yih}.
It should also be noticed that the equation for $\delta\widehat{\Sigma}$ decouples from the other equations
as no $\delta\widehat{\Sigma}$ terms appear in the equations for the other amplitudes and we could therefore elimate 
all $\partial_{\hat{t}}\delta\widehat{\Sigma}$ terms in Eqs.~(\ref{eqs:allk0}) by substituting the right-hand side of Eq.~(\ref{eq:Sk0}) for $\partial_{\hat{t}}\delta\widehat{\Sigma}$ in first five equations of (\ref{eqs:allk0}). Thus we only need to focus on Eqs.~(\ref{eq:Ak0}-\ref{eq:g1k0}).

By subtracting Eqs.~(\ref{eq:g0k0}) from (\ref{eq:Ak0}), we can find an equation expressed entirely in terms of $\alpha_n$, and 
subtracting Eqs.~(\ref{eq:g1k0}) from (\ref{eq:Dk0}) gives an equation in terms of $\delta_n$:
\begin{subequations}
\begin{align}
&\frac{d\alpha_m}{d\hat{t}}-8\sum_{n=0}^{\infty}\frac{d\alpha_n}{d\hat{t}}\frac{1}{(2n+1)^2\pi^2}=-A_c\left[\alpha_m-8\sum_{n=0}^{\infty}\frac{\alpha_n}{(2n+1)^2\pi^2}\right]\label{eq:Ael1},\\
&\frac{d\delta_m}{d\hat{t}}-8\sum_{n=0}^{\infty}\frac{d\delta_n}{d\hat{t}}\frac{1}{(2n+1)^2\pi^2}=-\left[\delta_m-8\sum_{n=0}^{\infty}\frac{\delta_n}{(2n+1)^2\pi^2}\right]\label{eq:Del1}.
\end{align}
\end{subequations}
The eigenvalues of Eq.~(\ref{eq:Ael1}) are $\lambda=A_c$ and of Eq.~(\ref{eq:Del1}) $\lambda=1$, reflecting the decay
of these amplitudes in time. When we next consider Eqs.~(\ref{eq:govg0}) and (\ref{eq:govg1}) and Eq.~(\ref{eq:Sk0} to eliminate
the $\frac{d\delta\widehat{\Sigma}}{d\hat{t}}$ terms, we find that the perturbations in the interface shear stresses obey 
\begin{subequations}
\begin{align}
&\frac{d\delta\hat{\dot{\gamma}}_0}{d\hat{t}}=-A_c\left(1-\frac{\hat{y}_0+\hat{y}_1}{2}\right)\delta\hat{\dot{\gamma}}_0+\left(1-\frac{\hat{y}_0+\hat{y}_1}{2}\right)\delta\hat{\dot{\gamma}}_1\label{eq:gel0},\\
&\frac{d\hat{\dot{\gamma}}_1}{d\hat{t}}=-\left(\frac{\hat{y}_0+\hat{y}_1}{2}\right)\hat{\dot{\gamma}}_1+A_c\left(\frac{\hat{y}_0+\hat{y}_1}{2}\right)\hat{\dot{\gamma}}_0\label{eq:gel1}.
\end{align}
\end{subequations}
Using Eq.~(\ref{eq:dg0dg1}) with the amplitudes $A_n$ and $D_n$ set to zero, as we have previously found that these decay in time,
and neglecting terms of order $\hat{y}_1-\hat{y}_0$  we readily find that Eqs.~(\ref{eq:gel0}) and (\ref{eq:gel1}) reduce to a single differential equation for $\hat{\dot{\gamma}}_0$ that reads
\begin{align}
\label{eq:g0red}
\frac{d\delta\hat{\dot{\gamma}}_0}{d\hat{t}}=-A_c\left(1-\hat{y}_0\right)\delta\hat{\dot{\gamma}}_0-\hat{y}_0\delta\hat{\dot{\gamma}}_0,
\end{align}
which clearly corresponds to a solution that decreases in time. As in this approximation $\delta\hat{\dot{\gamma}}_1$ is
proportional to $\delta\hat{\dot{\gamma}}_0$ the same holds true for $\delta\hat{\dot{\gamma}}_1$, showing that the the position of the interface is linearly stable.
This means that the stability of Eqs.~(\ref{eqs:allk0}) only depends on the  properties of the amplitudes $\beta_m$.
The even and odd modes of the amplitudes $\beta_m$ 
satisfy different evolution equations. 
Using the fact that $\alpha_n$, $\delta_n$ and $\delta\hat{\dot{\gamma}}_0$, $\delta\hat{\dot{\gamma}}_1$ decay in time, the stability properties of the amplitudes $\beta_n$ for both odd and even $n$ are determined by the equation
\begin{align}\label{eq:betaDC}
 \frac{d\beta_{n}}{d\hat{t}}=z\left(1-\frac{n^2\pi^2}{((2M+1)\pi-h)^2}\right)\beta_{n},
\end{align}
where $M$ is the band index. The system of equations (\ref{eq:betaDC}) gives eigenvalues  
\begin{align}
\label{eq:evB}
\lambda_{n}=-z\left(1-\frac{n^2\pi^2}{((2M+1)\pi-h)^2}\right). 
\end{align}
As the smallest eigenvalue is $\lambda_1$, the sign of this quantity determines the stability of the  amplitude
of $\beta_n$ and therefore the stability of the system. For $M=0$,
$\lambda_1>0$, which \emph{proves} that \emph{the two banded solution is neutrally stable}: all amplitudes $\alpha_n$, $\beta_n$, $\delta_n$, $\delta\hat{\dot{\gamma}}_0$, $\delta\hat{\dot{\gamma}}_1$  are decaying in time and $\alpha\delta\widehat{\Sigma}$ is staying constant in time. Multiple banded modes  with $M=1,2$ all have an eigenvalue $\lambda_1<0$, which indicates instability of these modes.

In conclusion we have proved that for small $\hat{D}$ the 
stationary shear band solution, with two bands is neutrally stable against non undulating perturbations
and all other stationary states are unstable.
Our analytical findings can be verified numerically. In Fig.~\ref{fig:fig5}   the results
of a numerical calculation of the 200 eigenvalues with smallest real part are displayed, for  $\hat{D}=10^{-5}$. Besides the zero eigenvalue corresponding to $\delta\widehat{\Sigma}$,
we indeed find  $\lambda=A_c$ corresponding to the amplitudes $\alpha_n$, and $\lambda=1$ corresponding to the amplitudes $\delta_n$, many times. The degeneracy is lifted by the small, but finite value of $\hat{D}$
and the finite size of the matrix. The eigenvalue $\lambda\,\approx\,0.1$ corresponds to $\beta_1$.
To check our result that all multiple bands solutions are unstable, we numerically calculated the smallest eigenvalues
of Eqs.~(\ref{eqs:allk0}) for the solutions $B$ and $C$, found in section \ref{sec:model}. We indeed find that solutions 
$B$ and $C$ are unstable to small nonundulating perturbations. This is shown in the  graph in Fig.~\ref{fig:figST}, where an eigenvalue with real part smaller than zero was found for both solutions $B$ and $C$.

We next turn to the general case in which the diffusion is no longer assumed to vanish and undulations in the 
perturbations are admitted. 

\subsection{Linear stability: nonzero diffusion}
\label{sec:Dnonzero}

The linear stability of the stationary solutions found in
section~\ref{sec:model} with respect to undulatory perturbations 
are examined numerically. We follow the same strategy as in the previous section,
projecting out the different sine and cosine modes, but this time keeping the imaginary 
parts of the evolution equations (\ref{eq:allds}) as $k$ is no longer presumed to vanish.
 

It again proves beneficial to use the scaled amplitudes $\alpha_n(k,\hat{t})$, $\beta_n(k,\hat{t})$, $\delta_n(k,\hat{t})$  defined in  Eq.~(\ref{eq:scaled}) rather  
than the amplitudes $a_n(k,\hat{t})$, $b_n(k,\hat{t})$, $d_n(k,\hat{t})$, for numerical convenience.
Once we have projected out the sine and cosine modes, we are left with the evolution
equations of the amplitudes. The details of this calculation are relegated to Appendix
\ref{sec:matderive}.  
Here we merely note that the general structure of the system of
differential equations, whose stability we would like to explore can easily be 
captured in a matrix equation.

By introducing a vector ${\vec
  {u}}=(\alpha_0,\cdots,\alpha_n,\beta_1,\cdots,\beta_n,\delta_0,\cdots,\delta_n,\delta\hat{\dot{\gamma}}_0,\delta\hat{\dot{\gamma}}_1,\delta\widehat{\Sigma})$, the matrix 
equivalent of the amplitude evolution is given by
\begin{equation}
\label{eq:matSL}
 {\bf S}\cdot\frac{d\vec{u}}{dt}=-({\bf L^R}-2ik{\bf L^I})\cdot{\vec {u}},
\end{equation}
where the $(3n+5)\times(3n+5)$ matrices ${\bf S}$ and ${\bf L^R}, {\bf L^I}$   can be easily read off from the amplitude equations (\ref{eq:govan}-\ref{eq:govdn}) in Appendix \ref{sec:matderive}. As in the previous section eigenvalues of the matrix 
${\bf S}^{-1}{\bf \tilde{L}}\,{\equiv}\,{\bf S}^{-1}{\bf L^R}-2ik{\bf S}^{-1}{\bf L^I}$ with
positive real part correspond to stable solutions. 
 The linear
stability of the stationary solutions found in section~\ref{subsec:d0}
with respect to non-undulatory perturbations is retrieved by setting $k=0$ in Eq,~(\ref{eq:matSL})

The results of the numerical calculations that were performed for this special case
are presented in Figs.~\ref{fig:fig5}  and \ref{fig:figST}  show good agreement with the 
analytical estimate for $\hat{D}\to 0$ and $k=0$.
The general case for $k{\neq}0$ does not allow any tractable analytical expressions for the eigenvalues.
We therefore have to resort to numerical methods. 


 \begin{figure}
   \includegraphics[width=9.70cm]{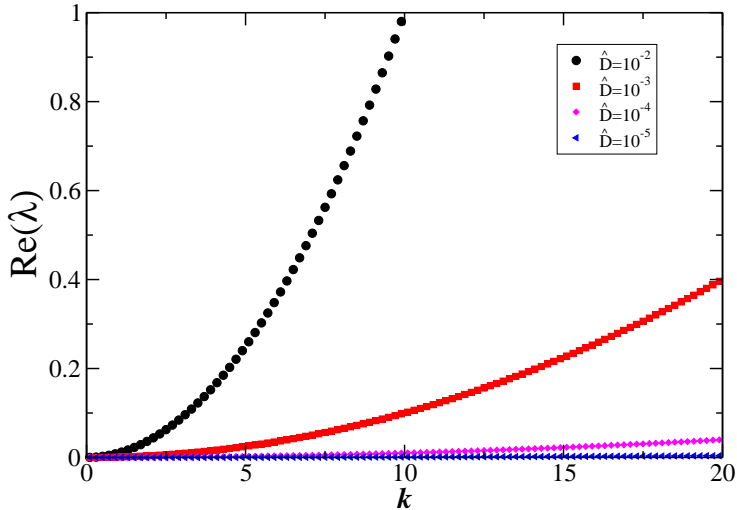}
   \caption{The eigenvalue with smallest real part
     as a function of $k$ for solution $A$
     using different values of the diffusion constant $\hat{D}$. The
     eigenvalue with smallest real part is observed to increase quadratically
     with  $k$. Moreover, the eigenvalues decrease with decreasing $\hat{D}$, eventually approaching zero,
a signal of  neutral stability.}
\label{fig:fig7}
 \end{figure}

To find the real part of eigenvalues for the stationary solution $A$, we diagonalize the matrix $\tilde{\bf L}$ numerically using the QR algorithm
from Ref.~\cite{numC}. We verified our results with the LAPACK
linear algebra package \cite{lapack}, and  verified convergence of the
eigenvalues up to $n=1000$. 

Figure \ref{fig:fig7} shows the real part
of the least stable eigenvalue, i.e. that with smallest real part, as a
function of $k$.  The dependence on $k$ is quadratic and for $k=0$ the smallest eigenvalue
is equal to 0. The stability decreases with decreasing $\hat{D}$;
for $\hat{D}$ decreasing, at a fixed value of $k=10$, from $\hat{D}=10^{-2}$ to $\hat{D}=10^{-5}$,
${\rm Re}(\lambda)$ changes from 1.0 to about $10^{-3}$.

 \begin{figure}
   \includegraphics[width=10.00cm]{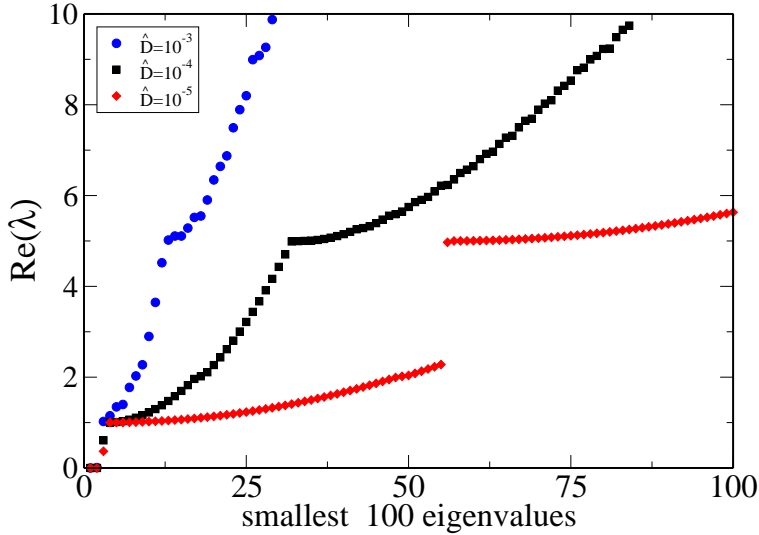}
   \caption{The distribution of the first 100 eigenvalues with smallest real part
     for fixed $k=3$ and  different values of the diffusion constant $\hat{D}$. The
     stair case is clearly recognizable and a clear accumulation of eigenvalues  occurs at ${\rm Re}(\lambda)=1$ and
${\rm Re}(\lambda)=A_c=5$.}
  
\label{fig:fig8}
 \end{figure}

To investigate the distribution of eigenvalues, we plot the 100 eigenvalues with the smallest real parts
for fixed $k=3$ in Fig.~\ref{fig:fig8}. One can still clearly distinguish the plateaus at ${\rm Re}(\lambda)=1$
and ${\rm Re}(\lambda)=A_c=5$. The plateaus become smaller for increasing $D$, as the eigenvalues 
increase more rapidly for larger $\hat{D}$. 

To verify our calculations, we have performed the same calculations for the smooth
 model originally introduced in Eq.~(\ref{eq:g}) (see Appendix
 \ref{sec:contmodel0} for details).  We find that the smooth model, indeed 
qualitatively reproduces the numerical results of the piecewise toy model.
\begin{figure}
   \includegraphics[width=16.00cm]{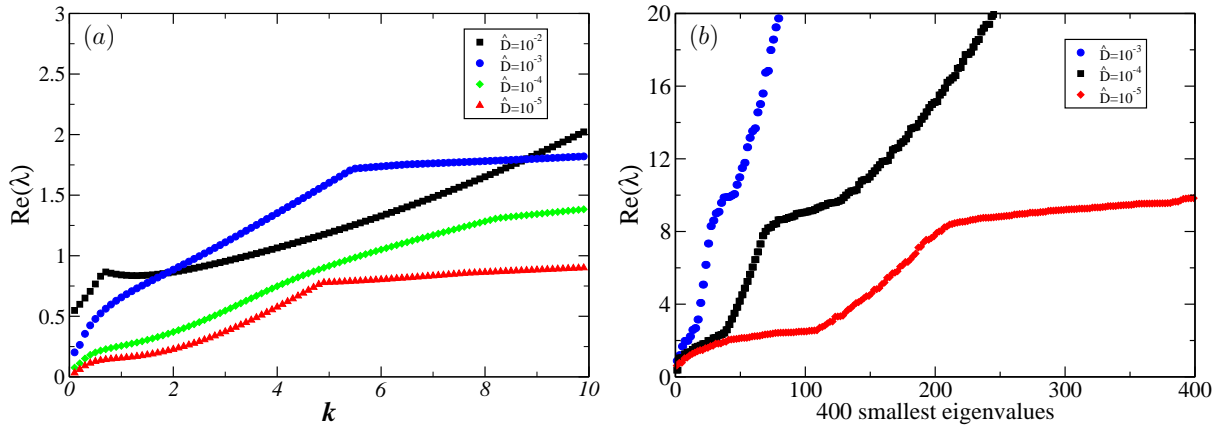}
   \caption{Real part of the eigenvalue with the smallest real part,
      as a function of the undulation wave
     number $k$ for four different values of the diffusion constant
     $\hat{D}$ for the \emph{continuous model} (a). For small $k<1$ the dependence of ${\rm Re}(\lambda)$ is
again approximately quadratic. For larger $k$ bumps occur, which are probably caused by the fact that the 
imaginary part of the equations start to dominate the eigenvalues. The distribution of  400 eigenvalues
with smallest real part (b) for $k=3$  is very similar to the distribution obtained by the piecewise toy model of
Fig.~\ref{fig:fig8}.}  
\label{fig:fig9}
 \end{figure}

Moreover,   for this model the shear banding state with two bands
 is also linearly stable with respect to undulations, as witnessed by the positivity of the real parts
of the eigenvalues in Fig.~\ref{fig:fig9}. Although we report here only the results  $A_c=5$, we verified
linear stability of the two-band solution for a large number of values of $A_c\in[1,10]$. This agrees   with the picture that arose from our analytical study for the 
case of zero diffusion, where we found that one eigenvalue is equal to zero whereas the others are concentrated at $A_c$ and $1$. The effect of diffusion is to shorten the plateaus where eigenvalues have real part $1$ or $A_c$, and this behavior persists for the entire range of $A_c$ studied.

The  continuous (Fig.~\ref{fig:fig9}(a)) and piecewise (Fig.~\ref{fig:fig7}) models display qualitatively
 similar features: the curves initially increase roughly quadratically  with
 $k$ and are always positive. For larger $k$ values we find ${\rm Re}(\lambda)\sim k$, 
which suggests that for this $k$ regime the imaginary components
of the evolution of the amplitudes dominate; and for  still larger values of $k$ the dependence of 
${\rm Re}(\lambda)(k)$ levels  off. This feature was occurs in the piecewise model at  much larger $k$-values not shown in Fig.~\ref{fig:fig7}.  In both cases smaller  $\hat{D}$ gives rise to  less stable eigenvalues. Precise comparison 
between the two models is difficult as the wavenumber $k$ is scaled 
with a factor $\alpha=20$ in the continuous case. This parameter $\alpha$ is necessary in order to find 
a stable shear banding solution in the continuous case \cite{SYC96}.
One striking difference between the curves in Fig.~\ref{fig:fig7} and Fig.~\ref{fig:fig9}(a) is the behavior in the limit $k\to 0$. In the smooth  model with $k\to 0$ the smallest eigenvalue depends on $\hat{D}$. The eigenvalue 0 results only when $\hat{D}\to 0$, in contrast to the piecewise model which is always neutrally stable with respect to non-undulatory perturbations. This
can mathematically be understood from the fact that the interface width scales as $\sqrt{\hat{D}}$ for  $\hat{D}\to 0$. Therefore the matching at the interface can be done in a smoother fashion when $\hat{D}$ becomes increasingly smaller. In the limit $\hat{D}\to0$ this will result in a solution to the piecewise model which approximates
the solution to the smooth model.      
If we compare Fig.~\ref{fig:fig9}(b)
with Fig.~\ref{fig:fig8}, we see the same qualitative behavior. Again plateaus occur, but because of the smoothness of the model 
the curves are more regular than those obtained for the piecewise toy  model. Nevertheless, Fig.~\ref{fig:fig9} suggests that the results for the piecewise model are generic and also valid for general smooth models
without normal stresses.

 From our findings we therefore
 conclude that normal stresses are generically responsible for
 rendering the Johnson-Segalman model linearly unstable for long
 wavelength undulations. This confirms the results which were first
 conjectured in Refs.~\cite{fielding04,WilsonFielding2006}.

\section{Discussion}
\label{sec:discuss}
We have studied, to a large extent analytically, a toy model for shear
banding without normal stresses, but with spatial gradient terms
(stress diffusion).  We captured the general characteristics of a
nonmonotonic stress-shear rate relation by introducing a piecewise
linear stress-shear rate curve, which makes analytical
calculations tractable. The results obtained for the piecewise model
were shown to be in qualitatively agreement from those obtained using
a smooth function $g$.

For the set of parameters chosen, we found multiple stationary
solutions when $\hat{D}<10^{-2}$. For $\hat{D}=10^{-3}$ we obtain
three stationary states: one being the commonly observed two-band
profile and the other two having three and four bands.  The
two-band profile was shown to be linearly stable with respect to
two-dimensional undulations. The linear stability of the two-band
profile is at variance with analogous results for Johnson-Segalman
(JS) model, which was shown by Fielding to have linear instabilities for a certain
range of $k$ vectors \cite{SMF2005} with $k\,\approx\,1$. This
strongly suggests that normal stresses, absent in our model, are
responsible for the linear instabilities arising in the JS model.
Related recent work by Fielding and co-workers
\cite{fielding06,WilsonFielding2006} showed that these linear
instabilities are suppressed if nonlinear effects are taken into
account. It might be that nonlinear perturbations induce instabilities
in the scalar model studied here. The behavior of such (weakly)
nonlinear instabilities is a subject of interest and future research.
\section*{Acknowledgements}
  It is a pleasure to thank the British Council and NWO for financial
  assistance. This work is part of the research programme of the
  `Stichting voor Fundamenteel Onderzoek der Materie (FOM)', which is
  financially supported by the `Nederlandse Organisatie voor
  Wetenschappelijk Onderzoek (NWO)'.

\appendix
\section{Integration Constants}
\label{sec:intconst}
The integration constants $c_1$,$c_2$, $d_1$, $d_2$ can be expressed in
terms of $\hat{y}_0$, the spatial position at which
$\hat{\dot{\gamma}}$ equals $\hat{\dot{\gamma}}_0$. From the four
equations (\ref{eq:c1}-\ref{eq:c2}) we find  
\begin{subequations}\label{eq:constants}
\begin{align}
  c_1&=\frac{\hat{\dot{\gamma}}_0- \frac{\alpha\widehat{\Sigma}}{A_c}}
  {\cosh\left(\sqrt{\frac{A_c}{\hat{D}}}\hat{y}_0\right)}\label{eq:ap1} \\
  c_2&={\frac{\hat{\dot{\gamma}}_1-\alpha\widehat{\Sigma}}
  {\cosh\left[(\hat{y}_1-1)/\sqrt{\hat{D}}\right]}} \label{eq:ap2}\\
  d_1&=\frac{(\alpha\widehat{\Sigma}- A_c \hat{\dot{\gamma}}_0)
  \cos\left(\sqrt{\frac{z}{\hat{D}}}\hat{y}_0\right)}{z} -
  \frac{\tanh\left(\sqrt{\frac{A_c}{\hat{D}}}\hat{y}_0\right)
  \sin\left(\sqrt{\frac{z}{\hat{D}}}\hat{y}_0\right)(A_c\hat{\dot{\gamma}}_0-
  \alpha\widehat{\Sigma})}{\sqrt{A_c z}}\\
  d_2&=\frac{(\alpha\widehat{\Sigma}- A_c    \hat{\dot{\gamma}}_0)
  \sin\left(\sqrt{\frac{z}{\hat{D}}}\hat{y}_0\right)}{z}+
  \frac{\tanh\left(\sqrt{\frac{A_c}{\hat{D}}}\hat{y}_0\right)
  \cos\left(\sqrt{\frac{z}{\hat{D}}}
  \hat{y}_0\right)(A_c\hat{\dot{\gamma}}_0-\alpha\widehat{\Sigma})}{ \sqrt{A_c z}},
\end{align}
\end{subequations}
where
$z=\frac{A_c\hat{\dot{\gamma}}_0-\hat{\dot{\gamma}}_1}{\hat{\dot{\gamma}}_1
  - \hat{\dot{\gamma}}_0}$.  From these equations, together with
Eqs.~(\ref{eq:c3}-\ref{eq:c4}), the values of $\hat{y}_0$ and
$\hat{y}_1$ can in principle be determined as a function of
$\widehat{\Sigma}$. To fix the value of the selected stress
$\widehat{\Sigma}^{\ast}$ an additional equation is required.  From
the global constraint of an imposed average shear rate
$\langle\hat{\dot{\gamma}}\rangle$ we can find an expression for
$\widehat{\Sigma}^{\ast}$. We substitute
Equations~(\ref{eq:constants}) for the constants $c_{1,2}$ and
$d_{1,2}$ into Eqs.~(\ref{eq:w1}-\ref{eq:w3}), and calculate the
average value of the shear rate directly by splitting up the integral
in three parts: $[0,\hat{y}_0]$,$[\hat{y}_0,\hat{y}_1]$,
$[\hat{y}_1,1]$. By equating the sum of the three parts to
$\langle{\hat{\dot{\gamma}}}\rangle$, we obtain the following
expression for $\widehat{\Sigma}$ in terms of $\hat{y}_0,\hat{y}_1$,
$\langle{\hat{\dot{\gamma}}}\rangle$, and
$\hat{\dot{\gamma}}_0,\hat{\dot{\gamma}}_1$:
\begin{align}
\alpha\widehat{\Sigma}^{\ast}=\frac{A_c\left\{\hat{\dot{\gamma}}_0(A_c+z)(\hat{y}_0-\hat{y}_1) -
    \sqrt{\hat{D}}\left[\hat{\dot{\gamma}}_0 
      \tanh\left(\sqrt{\frac{A_c}{\hat{D} }} \hat{y}_0\right) \left(\frac{{A_c}+z}{\sqrt{A_c}}\right)+
      \hat{\dot{\gamma}}_1 \tanh\left(\frac{1-\hat{y}_1}{\sqrt{\hat{D}}} \right)
    \left(z+1\right)\right] +{\langle{\hat{\dot{\gamma}}}\rangle} z\right\}}  {A_c(\hat{y}_0-\hat{y}_1+z-z\hat{y}_1) + \hat{y}_0 z -
 \sqrt{\hat{D}} \left[ \tanh\left(\sqrt{\frac{A_c}{\hat{D}}}\hat{y}_0\right) \left(\frac{{A_c} +z}{\sqrt{A_c}}\right)+
    A_c (1 +z)\tanh\left(\frac{1-\hat{y}_1}{\sqrt{\hat{D}}}\right)\right] }\label{eq:ap3}.
\end{align}

\section{Amplitude evolution equations}\label{sec:matderive}
The time evolution of the amplitudes $\alpha_n(k,\hat{t})$ and $\delta\hat{\dot{\gamma}}_0(k,\hat{t})$
can be determined by substituting $\delta\hat{\dot{\gamma}}^{I}(\hat{x},\hat{y},\hat{t})$ in
Eq.~(\ref{seq:dtsigma1}). When we project out the cosine modes and use the definition of the scaled amplitudes
$\alpha_n(k,\hat{t})=a_n(k,\hat{t})(2n+1)\pi(-1)^n/4$, we obtain

\begin{align}
\label{eq:govan}
&\frac{d\alpha_m}{d\hat{t}}-\left(\frac{\partial\hat{\dot{\gamma}}^{1D}}{\partial\hat{y}}\right)_{\hat{y}_0}\partial_{\hat{t}}\delta \hat{y}_0(k,\hat{t})-\partial_{\hat{t}}\alpha\delta\widehat{\Sigma}(k,\hat{t})= (A_c+\hat{D}k^2)\left(\frac{\partial\hat{\dot{\gamma}}^{1D}}{\partial\hat{y}}\right)_{\hat{y}_0}\delta \hat{y}_0(k,\hat{t})-\left(A_c+\hat{D}k^2+\hat{D}\left(\frac{\pi(2m+1)}{2\hat{y}_0}\right)^2\right)\alpha_m\nonumber\\
&+\hat{D}k^2\alpha\widehat{\Sigma}(k,\hat{t})-\frac{ik(-1)^m(2m+1)\pi}{2\hat{y}_0}\left[\sum_{n=1}^{\infty}\alpha_n(-1)^n\frac{16\hat{y}_0^2}{(2n+1)^3\pi^3}\left[R^A_{mn}-T^A_m\right] +\left(\frac{\partial\hat{\dot{\gamma}}^{1D}}{\partial\hat{y}_0}\right)_{\hat{y}_0}\delta\hat{y}_0(k,\hat{t}) F^A_m\right.\nonumber\\
&+\left.\sum_{n=1}^{\infty}\left[L^A_{mn}+\frac{D}{A_c}R^A_{mn}\right]\frac{4\alpha_n(-1)^n}{(2n+1)\pi} -\left(J^A_m+\frac{\hat{D}}{A_c}T^A_m\right)\left[\alpha\delta\widehat{\Sigma}+\left(\frac{\partial\hat{\dot{\gamma}}^{1D}}{\partial\hat{y}}\right)_{\hat{y}_0}\delta\hat{y}_0(k,\hat{t})\right]\right].
\end{align}

\begin{align}
\label{eq:govg0}
&8\sum_{n=0}^{\infty}\frac{d\alpha_n}{d\hat{t}}\frac{1}{(2n+1)^2\pi^2}-\alpha\partial_{\hat{t}}\delta\widehat{\Sigma}+\partial_{\hat{t}}\delta\hat{\gamma}_0=\hat{D}k^2\alpha\delta\widehat{\Sigma}-(A_c+\hat{D}k^2)\delta\hat{\gamma}_0-8(A_c+\hat{D}k^2)\sum_{n=0}^{\infty}\frac{\alpha_n}{(2n+1)^2\pi^2}-\frac{2\hat{D}}{\hat{y}_0^2}\sum_{n=0}^{\infty}\alpha_n\nonumber\\
&-\frac{ik}{\hat{y}_0}\left\{-\sum_{n=0}^{\infty}\frac{\alpha_n(1-(-1)^n)16\hat{y}_0^2}{(2n+1)^3\pi^3}\hat{\dot{\gamma}}_0
+\sum_{n=0}^{\infty}\frac{8\hat{y}_0\alpha_n}{(2n+1)^2\pi^2}\left[\frac{\alpha\widehat{\Sigma}\hat{y}_0}{A_c}+(\hat{\dot{\gamma}}_0-\frac{\alpha\widehat{\Sigma}}{A_c})\sqrt{\hat{D}}\right]\right.\nonumber\\
&-\left.\alpha\delta\widehat{\Sigma}\left[\frac{\alpha\widehat{\Sigma} \hat{y}_0^2}{2A_c}+(\hat{\dot{\gamma}}_0-\frac{\alpha\widehat{\Sigma}}{A_c})\frac{\hat{D}}{A_c}\right]+\hat{y}_0\delta\gamma_0\left(\frac{\alpha\widehat{\Sigma}\hat{y}_0}{A_c}+\left(\hat{\dot{\gamma}}_0-\frac{\alpha\widehat{\Sigma}}{A_c}\right)\sqrt{\hat{D}}\right)-\frac{\hat{y}_0^2\hat{\dot{\gamma}}_0\delta\gamma_0}{2}\right\}\nonumber\\
&\equiv\hat{D}k^2\alpha\delta\widehat{\Sigma}-(A_c+\hat{D}k^2)\delta\hat{\gamma}_0-8(A_c+\hat{D}k^2)\sum_{n=0}^{\infty}\frac{\alpha_n}{(2n+1)^2\pi^2}-\frac{2\hat{D}}{\hat{y}_0^2}\sum_{n=0}^{\infty}\alpha_n-\frac{ik}{\hat{y}_0}\Pi_1(k,\hat{t}).
\end{align}
In Eq.~(\ref{eq:govg0}) we defined the real function $\Pi_1(k,\hat{t})$ as the imaginary contribution to the evolution 
of $\delta\hat{\gamma}_0$. 
The coefficients $R_{mn}^A,T_m^A,F^A_m,L_{mn}^A,J^A_m$ in Eqs.~(\ref{eq:govan}) are
defined in terms of integrals of functions on the interval $[0,\hat{y}_0]$:
\begin{subequations}
    \label{eq:integralcoef}
  \begin{align}
    R_{mn}^A &=\int_0^{\hat{y}_0} \cos\left(\frac{(2m+1)\pi\hat{y}}{2\hat{y}_0}\right)\cos\left(\frac{(2n+1)\pi\hat{y}}{2\hat{y}_0}\right)\frac{\partial\hat{\dot{\gamma}}^{1D}}{\partial\hat{y}}\,d\hat{y}\\
    F^A_{m} & =\int_0^{\hat{y}_0} \cos\left(\frac{(2m+1)\pi\hat{y}}{2\hat{y}_0}\right)\frac{\hat{y}^2}{2}\frac{\partial\hat{\dot{\gamma}}^{1D}}{\partial\hat{y}}\,d\hat{y}\\
    L_{mn}^A & =\int_0^{\hat{y}_0} \cos\left(\frac{(2m+1)\pi\hat{y}}{2\hat{y}_0}\right)\cos\left(\frac{(2n+1)\pi\hat{y}}{2\hat{y}_0}\right)\frac{\alpha\widehat{\Sigma}\hat{y}}{A_c}\,d\hat{y},
  \end{align}
\end{subequations}
and $T_m^A=R_{m,-\frac{1}{2}}^A$, and  $J^A_m=L_{m,-\frac{1}{2}}^A$.
The evolution equations for the amplitudes $\delta_n(k,\hat{t})$ is derived
by substituting the expression for
$\delta\hat{\dot{\gamma}}^{III}(\hat{x},\hat{y},\hat{t})$ in
Eq.~(\ref{seq:dtsigma3}), projecting out the cosine mode,s and using the scaling relation between $d_n$ and $\delta_n$ , yielding
\begin{align}
\label{eq:govdn}
&\frac{d\delta_m}{d\hat{t}}-\left(\frac{\partial\hat{\dot{\gamma}}^{1D}}{\partial\hat{y}}\right)_{\hat{y}_1}\partial_{\hat{t}}\delta \hat{y}_1(k,\hat{t})-\partial_{\hat{t}}\alpha\delta\widehat{\Sigma}(k,\hat{t})= (1+\hat{D}k^2)\left(\frac{\partial\hat{\dot{\gamma}}^{1D}}{\partial\hat{y}}\right)_{\hat{y}_1}\delta \hat{y}_1(k,\hat{t})-\left(1+\hat{D}k^2+\hat{D}\left(\frac{\pi(2m+1)}{2(1-\hat{y}_1)}\right)^2\right)\delta_m\nonumber\\
&+\hat{D}k^2\alpha\widehat{\Sigma}(k,\hat{t})-\frac{ik(-1)^m(2m+1)\pi}{2(1-\hat{y}_1)}\left[\sum_{n=1}^{\infty}\delta_n(-1)^n\frac{16(1-\hat{y}_1)^2}{(2n+1)^3\pi^3}\left[\tilde{R}^D_{mn}-\tilde{T}^D_m\right] +\left(\frac{\partial\hat{\dot{\gamma}}^{1D}}{\partial\hat{y}_1}\right)_{\hat{y}_1}\delta\hat{y}_1(k,\hat{t})\tilde{ F}^D_m\right.\nonumber\\
&+\left.\sum_{n=1}^{\infty}\left[\tilde{L}^D_{mn}+\hat{D}\tilde{R}^D_{mn}\right]\frac{4\delta_n(-1)^n}{(2n+1)\pi} -\left(\tilde{J}^D_m+\hat{D}\tilde{T}^D_m\right)\left[\alpha\widehat{\Sigma}+\left(\frac{\partial\hat{\dot{\gamma}}^{1D}}{\partial\hat{y}}\right)_{\hat{y}_1}\delta\hat{y}_1(k,\hat{t})\right]\right]-ik\langle\hat{\dot{\gamma}}\rangle\delta_m\nonumber\\
&+ik\left[\alpha\delta\widehat{\Sigma}+\left(\frac{\partial \hat{\dot{\gamma}}^{1D}}{\partial\hat{y}}\right)_{\hat{y}_1}\delta\hat{y}_1\right]\langle\hat{\dot{\gamma}}\rangle,
\end{align}
where the functions
$\tilde{R}_{mn}^D,\tilde{T}_m^D,\tilde{F}^D_m,\tilde{L}_{mn}^D,\tilde{J}^D_m$
are defined similarly to Eqs.~(\ref{eq:integralcoef}), with $\hat{y}_0$
everywhere replaced by $1-\hat{y}_1$ and $A_c$ set to 1.
If we integrate the amplitude equation for $d_n(k,t)$  from $\hat{y}_1$ to $1$, we obtain the evolution equation  for the $\delta\hat{\dot{\gamma}}_1(k,\hat{t})$,
\begin{align}
\label{eq:govg1}
&8\sum_{n=0}^{\infty}\frac{d\delta_n}{d\hat{t}}\frac{1}{(2n+1)^2\pi^2}-\alpha\partial_{\hat{t}}\delta\widehat{\Sigma}+\partial_{\hat{t}}\delta\hat{\gamma}_1=\hat{D}k^2\alpha\delta\widehat{\Sigma}-(1+\hat{D}k^2)\delta\hat{\gamma}_1-8(1+\hat{D}k^2)\sum_{n=0}^{\infty}\frac{\delta_n}{(2n+1)^2\pi^2}-\frac{2\hat{D}}{(1-\hat{y}_1)^2}\sum_{n=0}^{\infty}\delta_n\nonumber\\
&-\frac{ik}{(1-\hat{y}_1)}\left\{-\sum_{n=0}^{\infty}\frac{\delta_n(1-(-1)^n)16(1-\hat{y}_1)^2}{(2n+1)^3\pi^3}\hat{\dot{\gamma}}_1
+\sum_{n=0}^{\infty}\frac{8(1-\hat{y}_1)\delta_n}{(2n+1)^2\pi^2}\left[\langle\hat{\dot{\gamma}}\rangle+\alpha\widehat{\Sigma}(\hat{y}_1-1)-(\hat{\dot{\gamma}}_1-\alpha\widehat{\Sigma})\sqrt{\hat{D}}\right]\right.\nonumber\\
&-\alpha\delta\widehat{\Sigma}\left[\langle\hat{\dot{\gamma}}\rangle(1-\hat{y}_1)-\frac{\alpha\widehat{\Sigma} (1-\hat{y}_1)^2}{2}-(\hat{\dot{\gamma}}_1-\alpha\widehat{\Sigma})\hat{D}\right]-(1-\hat{y}_1)\delta\gamma_1\left(\langle\hat{\dot{\gamma}}\rangle-\alpha\widehat{\Sigma}(1-\hat{y}_1)+\left(\hat{\dot{\gamma}}_1-\alpha\widehat{\Sigma}\right)\sqrt{\hat{D}}\right)\nonumber\\
&+\left.\frac{(1-\hat{y}_1)^2\hat{\dot{\gamma}}_1\delta\gamma_1}{2}\right\}{\equiv}\hat{D}k^2\alpha\delta\widehat{\Sigma}-(1+\hat{D}k^2)\delta\hat{\gamma}_1-8(1+\hat{D}k^2)\sum_{n=0}^{\infty}\frac{\delta_n}{(2n+1)^2\pi^2}-\frac{2\hat{D}}{(1-\hat{y}_1)^2}\sum_{n=0}^{\infty}\delta_n-\frac{ik}{(1-\hat{y}_1)}\Pi_2(k,\hat{t}),
\end{align}
where $\Pi_2(k,\hat{t})$ is the imaginary contribution to the evolution of $\delta\hat{\gamma}_1$.
The equation for the amplitudes $b_n(k,\hat{t})$ can be found from the evolution 
equation of $q$ (Eq.~\ref{eq:qt}) and substituting expression (\ref{eq:defq}), which leads to a lengthy differential 
equation for $b_n(k,\hat{t})$.  After projecting out the sine modes over $[\hat{y}_0,\hat{y}_1]$, we find
\begin{align}\label{eq:govbn}
&\frac{d\beta_m}{dt}+\frac{1}{2}\left[(-1)^m\partial_t\left(\frac{\partial\hat{\dot{\gamma}}^{1D}}{\partial\hat{y}}\right)_{\hat{y}_1}\delta\hat{y}_1-\partial_t\left(\frac{\partial\hat{\dot{\gamma}}^{1D}}{\partial\hat{y}}\right)_{\hat{y}_0}\delta\hat{y}_0\right]-\frac{1}{2}\partial_t\alpha\delta\widehat{\Sigma}(1-(-1)^m)=\beta_m\left[z-\hat{D}k^2-\hat{D}\left(\frac{m\pi}{(2M+1)\pi-h}\right)^2\right]\nonumber\\
&+\frac{z-\hat{D}k^2}{2}\left[(-1)^m\left(\frac{\partial\hat{\dot{\gamma}}^{1D}}{\partial\hat{y}}\right)_{\hat{y}_1}\delta\hat{y}_1-\left(\frac{\partial\hat{\dot{\gamma}}^{1D}}{\partial\hat{y}}\right)_{\hat{y}_0}\delta\hat{y}_0\right]+\frac{1}{2}\hat{D}k^2\alpha\delta\widehat{\Sigma}(1-(-1)^m)\nonumber\\
&-\frac{ik m\pi}{2(\hat{y}_1-\hat{y}_0}\left\{-\left(\frac{\partial\hat{\dot{\gamma}}^{1D}}{\partial\hat{y}}\right)_{\hat{y}_0}\frac{\hat{y}_0^2\delta\hat{y}_0}{2}\left[-d_1 P_1(m)+d_2 P_2(m)\right]+\hat{y}_0\delta\hat{y}_0\left(\frac{\partial\hat{\dot{\gamma}}^{1D}}{\partial\hat{y}}\right)_{\hat{y}_0}[-d_1V_1(m)+d_2V_2(m)]\right.\nonumber\\
&+\sum_{n=0}^{\infty}\frac{16\alpha_n\hat{y}_0^2(1-(-1)^n)}{(2n+1)^3\pi^3}\left[-d_1P_1(m)+d_2P_2(m)\right]-\sum_{n=0}^{\infty}\frac{8\hat{y}_0\alpha_n}{(2n+1)^2\pi^2}\left[-d_1V_1(m)+d_2V_2(m)+d_1\hat{y}_0P_1(m)-d_2\hat{y}_0P_2(m)\right]\nonumber\\
&-\sum_{n=0}^{\infty}\frac{4\beta_n(\hat{y}_1-\hat{y}_0)}{n^2\pi^2}\left[-d_1V_1(m)+d_2V_2(m)+d_1\hat{y}_0P_1(m)-d_2\hat{y}_0P_2(m)\right]+\sum_{n=1}^{\infty}\left[\frac{4\beta_n(\hat{y}_1-\hat{y}_0)^2}{n^3\pi^3}-\frac{4\hat{D}\beta_n}{\pi nz}\right]\nonumber\\
&\times\left[-d_1G_1(m,n)+d_2G_2(m,n)\right]+\sum_{n=1}^{\infty}\frac{4\beta_n Q_1(m,n)}{n\pi}\tilde{C}_2+
\frac{-\hat{y}_0+(-1)^m\hat{y}_1}{m\pi}\left[\left(\frac{\partial\hat{\dot{\gamma}}^{1D}}{\partial\hat{y}}\right)_{\hat{y}_1}\delta\hat{y}_1-\left(\frac{\partial\hat{\dot{\gamma}}^{1D}}{\partial\hat{y}}\right)_{\hat{y}_0}\delta\hat{y}_0\right]\tilde{C}_1\nonumber\\
&+\frac{1-\hat{y}_0+(\hat{y}_1-1)(-1)^m\hat{y}_1}{m\pi}\left[\hat{y}_0\left(\frac{\partial\hat{\dot{\gamma}}^{1D}}{\partial\hat{y}}\right)_{\hat{y}_1}\delta\hat{y}_1-\hat{y}_1\left(\frac{\partial\hat{\dot{\gamma}}^{1D}}{\partial\hat{y}}\right)_{\hat{y}_0}\delta\hat{y}_0\right]\tilde{C}_1-\tilde{C}_2W(m)\left[\left(\frac{\partial\hat{\dot{\gamma}}^{1D}}{\partial\hat{y}}\right)_{\hat{y}_1}\delta\hat{y}_1-\left(\frac{\partial\hat{\dot{\gamma}}^{1D}}{\partial\hat{y}}\right)_{\hat{y}_0}\delta\hat{y}_0\right]\nonumber\\
&+\sqrt{\frac{\hat{D}}{z}}\left(\frac{1}{(2M+1)\pi-h}\right)\left[d_1V_1(m)-d_2V_2(m)\right]\left[\left(\frac{\partial\hat{\dot{\gamma}}^{1D}}{\partial\hat{y}}\right)_{\hat{y}_0}\delta\hat{y}_0-\left(\frac{\partial\hat{\dot{\gamma}}^{1D}}{\partial\hat{y}}\right)_{\hat{y}_1}\delta\hat{y}_1\right]\nonumber\\
&+\sqrt{\frac{\hat{D}}{z}}\left(\frac{1}{(2M+1)\pi-h}\right)\left[d_1P_1(m)-d_2P_2(m)\right]\left[\left(\frac{\partial\hat{\dot{\gamma}}^{1D}}{\partial\hat{y}}\right)_{\hat{y}_1}\hat{y}_0\delta\hat{y}_1-\left(\frac{\partial\hat{\dot{\gamma}}^{1D}}{\partial\hat{y}}\right)_{\hat{y}_0}\hat{y}_1\delta\hat{y}_0\right]-\alpha\delta\widehat{\Sigma}(1-(-1)^m)\frac{\hat{y}_1-\hat{y}_0}{m\pi}\tilde{C}_1\nonumber\\
&+\left.\alpha\tilde{C}_2(\hat{y}_1-\hat{y}_0)\delta\widehat{\Sigma}\frac{(-\hat{y}_0+\hat{y}_1(-1)^m)}{m\pi}-\alpha\delta\widehat{\Sigma}\left[d_1P_1(m)-d_2P_2(m)\right]\frac{\hat{D}}{z}\right\}-ik\tilde{C}_1\beta_m,
\end{align}
where 
\begin{subequations}
\begin{align}
&P_1(m)=\sqrt{\frac{z}{\hat{D}}}\int_{\hat{y}_0}^{\hat{y}_1}\sin\left(\frac{m\pi(\hat{y}-\hat{y}_0)}{\hat{y}_1-\hat{y}_0}\right)\,\sin\left(\sqrt{\frac{z}{\hat{D}}}\hat{y}\right)\,d\hat{y}\\
&V_1(m)= \sqrt{\frac{z}{\hat{D}}}\int_{\hat{y}_0}^{\hat{y}_1}\sin\left(\frac{m\pi(\hat{y}-\hat{y}_0)}{\hat{y}_1-\hat{y}_0}\right)\,\sin\left(\sqrt{\frac{z}{\hat{D}}}\hat{y}\right)\hat{y}\,d\hat{y}\\
&G_1(m,n)=\sqrt{\frac{z}{\hat{D}}}\int_{\hat{y}_0}^{\hat{y}_1}\sin\left(\frac{n\pi(\hat{y}-\hat{y}_0)}{\hat{y}_1-\hat{y}_0}\right)\sin\left(\frac{m\pi(\hat{y}-\hat{y}_0}{\hat{y}_1-\hat{y}_0}\right)\,\sin\left(\sqrt{\frac{z}{\hat{D}}}\hat{y}\right)\hat{y}\,d\hat{y}\\
&Q_1(m,n)=\int_{\hat{y}_0}^{\hat{y}_1}\sin\left(\frac{n\pi(\hat{y}-\hat{y}_0)}{\hat{y}_1-\hat{y}_0}\right)\sin\left(\frac{m\pi(\hat{y}-\hat{y}_0}{\hat{y}_1-\hat{y}_0)}\right)\hat{y}\,d\hat{y}\\
&W(m)=\int_{\hat{y}_0}^{\hat{y}_1}\sin\left(\frac{n\pi(\hat{y}-\hat{y}_0)}{\hat{y}_1-\hat{y}_0)}\right)\frac{\hat{y}^2}{\hat{y}_1-\hat{y}_0}\,d\hat{y}.
\end{align}
\end{subequations}
The functions $P_2(m)$, $V_2(m)$, $G_2(m,n)$ are obtained if the $\sin\left(\sqrt{\frac{z}{\hat{D}}}\hat{y}\right)$ function is replaced by $\cos\left(\sqrt{\frac{z}{\hat{D}}}\hat{y}\right)$
The constants $\tilde{C}_1$ and $\tilde{C}_2$ are defined by
\begin{subequations}
\begin{align}
\tilde{C}_1&=\frac{\alpha\widehat{\Sigma}-A_c\hat{\dot{\gamma}}_0}{z A_c}\hat{y}_0(z+A_c)+\sqrt{\frac{\hat{D}}{A_c}}\left(
\hat{\dot{\gamma}}_0-\frac{\alpha\widehat{\Sigma}}{A_c}\right)-\sqrt{\frac{\hat{D}}{z}}\left[d_1\sin\left(\sqrt{\frac{z}{\hat{D}}}\hat{y}_0\right)-d_2\cos\left(\sqrt{\frac{z}{\hat{D}}}\hat{y}_0\right)\right],\\
\tilde{C}_2&=\hat{\dot{\gamma}}_0+\frac{A_c\hat{\dot{\gamma}}_0-\alpha\widehat{\Sigma}}{z}. 
\end{align}
\end{subequations}
The final evolution equation we need is that for $\delta\widehat{\Sigma}$, which is obtained
by using Eqs.~(\ref{eq:govg0}) and (\ref{eq:govg1})  and substituting the expressions for $\partial_t\delta\hat{\gamma}_0$, $\partial_t\delta\hat{\gamma}_1$ in the differentiated Eq.~(\ref{eq:dg0dg1}). We then find that the evolution equations for the average stress perturbation $\delta\widehat{\Sigma}$ read:
\begin{align}
\label{eq:govsig}
&\alpha\partial_{\hat{t}}\delta\widehat{\Sigma}-4(\hat{y}_1-\hat{y}_0)\sum_{n=0}^{\infty}\frac{d\alpha_n}{dt}\frac{1}{(2n+1)^2\pi^2}-4(\hat{y}_1-\hat{y}_0) \sum_{n=0}^{\infty}\frac{d\delta_n}{dt}\frac{1}{(2n+1)^2\pi^2}+8(\hat{y}_1-\hat{y}_0)\sum_{n=0}^{\infty}\frac{d\beta_{2n+1}}{d\hat{t}}\frac{1}{(2n+1)^2\pi^2}=\nonumber\\
&+4(1+\hat{D}k^2)(2-\hat{y}_1-\hat{y}_0)\sum_{n=0}^{\infty}\frac{\delta_n}{(2n+1)^2\pi^2}+
4(A_c+\hat{D}k^2)(\hat{y}_1+\hat{y}_0)\sum_{n=0}^{\infty}\frac{\alpha_n}{(2n+1)^2\pi^2}+\frac{\hat{D}(\hat{y}_0+\hat{y}_1)}{\hat{y}_0^2}\sum_{n=0}^{\infty}\frac{\alpha_n}{(2n+1)^2\pi^2}\nonumber\\
&+\frac{\hat{D}(2-\hat{y}_0-\hat{y}_1)}{(1-\hat{y}_1)^2}\sum_{n=0}^{\infty}\frac{\delta_n}{(2n+1)^2\pi^2}+(1+\hat{D}k^2)(2-\hat{y}_1-\hat{y}_0)\frac{\delta\hat{\gamma}_1}{2}+(A_c+\hat{D}k^2)(\hat{y}_1+\hat{y}_0)\frac{\delta\hat{\gamma}_0}{2}-\alpha\hat{D}k^2\delta\widehat{\Sigma}\nonumber\\
&+ik\left(\frac{(\hat{y}_0+\hat{y}_1)}{2\hat{y}_0}\Pi_1(k,\hat{t})+\frac{2-\hat{y}_1-\hat{y}_0}{2(1-\hat{y}_1)}\Pi_2(k,\hat{t})\right).
\end{align}

 
\section{Continuous model}\label{sec:contmodel0}
In this appendix we present the results of stability analysis of the
first model, in which the nonmonotonic stress is encoded in the smooth
function $g(\xi)$ (Eq.~\ref{eq:g}). A straightforward calculation
shows that for this case the amplitudes $a_n(k,\hat{t})$ are not given by
Eq.~(\ref{eq:govan}), but instead obey
\begin{align}
\label{eq:lambdacont}
\sum_{n=0}^{\infty}\frac{da_n(k,\hat{t})}{d\hat{t}}\cos(n\pi
\hat{y})&=-\sum_{n=0}^{\infty}\left[1+\hat{D}(n^2\pi^2+
  k^2)\right]a_n(k,\hat{t})\cos(n\pi \hat{y})
-\alpha g'[\alpha(\langle \hat{\sigma}_p^{1D}\rangle+\langle{\hat{\dot{\gamma}}}\rangle/\alpha-\hat{\sigma}_p^{1D}(y))]
\sum_{n=1}^{\infty}a_n(k,\hat{t})\cos(n\pi \hat{y})\\
&+ik
\left\{\alpha\sum_{n=1}^{\infty}a_n(k,\hat{t})\left[\frac{\cos(n\pi
      \hat{y})-1}{n^2\pi^2} \right]\partial_{\hat{y}}  \hat{\sigma}_p^{1D}(\hat{y}) -
  \sum_{n=0}^{\infty}a_n(k,\hat{t})\hat{v}^{1D}(\hat{y})\cos(n\pi
  \hat{y})\right\}.\nonumber
\end{align}
Projecting out the cosines, we have the following matrix equation
\begin{equation}
\dot{\bf a}=-\left({\bf P}^{cont}\,{\cdot}\,{\bf a}-2i\alpha k\, {\bf
    Q}^{cont}\,{\cdot}\,{\bf a}\right),
\end{equation}
where the dot denotes denotes differentiation with respect to time.
The entries of matrix ${\bf P}^{cont}$ are given as
\begin{align}
P_{00}^{cont}&=(1+\hat{D} k^2)\\
P_{0n}^{cont}&=\alpha\int_0^1 g'[\alpha(\langle
\hat{\sigma}_p^{1D}\rangle + \langle{\hat{\dot{\gamma}}}\rangle/\alpha-
\hat{\sigma}_p^{1D}(y))] \cos(n\pi y)\,dy,\,\,(n\,{\ge}\,1)\\
P_{n0}^{cont}&=0,\,\,(n\,{\ge}\,1)\\
P_{mn}^{cont}&=(1+\hat{D}(n^2\pi+
k^2))\delta_{nm}+2\alpha\int_0^1 g'[\alpha(\langle
\hat{\sigma}_p^{1D}\rangle  + \langle{\hat{\dot{\gamma}}}\rangle/\alpha
-\hat{\sigma}_p^{1D}(y))] \cos(n\pi y)\cos(m\pi y)\,dy,\,\,\,(m,n\,{\ge}\,1).
\end{align}
For the matrix ${\bf Q}^{cont}$, we find the entries
\begin{align}
Q_{00}^{cont}&=-\frac{\langle{\hat{\dot{\gamma}}} \rangle}{4\alpha}+
\frac{\langle\hat{\sigma}_p^{1D}\rangle}{4}- \frac{1}{2}\int_0^1
y\hat{\sigma}_p^{1D}(y)\,dy\\
Q_{0n}^{cont}&=
\begin{cases}
  \frac{\langle\hat{\sigma}_p^{1D}\rangle+\frac{\langle{\hat{\dot{\gamma}}}\rangle}{\alpha}-
\hat{\sigma}^{1D}_p(1)}{n^2\pi^2}
 &  (\textrm{$n$ odd}) \\
0 &  (\textrm{$n$ even})
\end{cases}\\
Q_{n0}^{cont}&=
\begin{cases}
      \frac{2(\langle\hat{\sigma}_p^{1D}\rangle
        +\frac{\langle{\hat{\dot{\gamma}}}\rangle}{\alpha})}{n^2\pi^2}-
        \int_0^1\hat{\sigma}_p^{1D}(y)\frac{\sin(n\pi y)}{n\pi}\,dy &
        (\textrm{$n$ odd})\\
- \int_0^1\hat{\sigma}_p^{1D}(y)\frac{\sin(n\pi y)}{n\pi}\,dy
        &(\textrm{$n$ even})
\end{cases}\\
Q_{nn}^{cont}&=-\frac{1}{4}\left(\frac{\langle{{\gamma}}\rangle}{\alpha}+
  \langle \hat{\sigma}_p^{1D}\rangle\right)
  +\frac{3}{4n\pi}\int_0^1\hat{\sigma}_p^{1D}(y)\sin(2\pi n
  y)\,dy\nonumber\\
& -\frac{1}{2}\int_0^1\hat{\sigma}_p^{1D}(y)y\,dy-
  \frac{1}{n\pi}\int_0^1\sin(n\pi y)\hat{\sigma}_p^{1D}(y)\,dy+
  \left(\frac{1-(-1)^n}{n^2\pi^2}\right)\hat{\sigma}_p^{1D}(1)\\
Q_{mn}^{cont}&=-\left(\frac{
  \langle{\hat{\dot{\gamma}}}\rangle}{\alpha}+ \langle
  \hat{\sigma}_p^{1D}\rangle\right)\left(\frac{-2m^2-2n^2+
  (-1)^{m-n}(m+n)^2+ (-1)^{m+n}(m-n)^2}{2\pi^2(m-n)^2(m+n)^2}\right),
  \,\,\,(n\,{\ge}\,1)\nonumber\\
& -\int_0^1\frac{\hat{\sigma}_p^{1D}(y)\left[m\sin((n-m)\pi y)+
  n\sin((n-m)\pi y)-m\sin((n+m)\pi y)+n\sin((n+m)\pi
  y)\right]}{2\pi^2(n^2-m^2)}\nonumber\\
& +\int_0^1\left[\frac{\sin(n\pi y)\cos(m\pi y)}{n\pi}\right]
  \hat{\sigma}_p^{1D}(y)\,dy+(-1)^m\hat{\sigma}_p^{1D}(1)\left[\frac{(-1)^n-1}{n^2\pi^2}\right]\nonumber\\
& +\int_0^1\hat{\sigma}_p^{1D}(y)\left[\frac{\cos(n\pi
    y)-1}{n^2\pi^2}\right]m\pi\,\sin(m\pi y)\,dy ,\qquad(m{\neq}n,m,n\,{\ge}\,1)
\end{align}
All the matrix elements have to be computed numerically, after which
the linear stability of the solutions can be examined along the same
lines as for the piecewise model. From these computations, we
conclude  that the continuous equivalent of the piecewise
system exhibits the same stability behavior as the piecewise model: it
is  linearly stable. For long wavelength undulations the dispersion
relation is again dominated by the real diagonal matrix elements. That is, we
find again a quadratic dependence of the eigenvalue with the smallest real part on $k$.

\end{onecolumn}


\begin{thebibliography}{10}

\bibitem{cohen2006sya}
I. Cohen, B. Davidovitch, A. Schofield, M. Brenner, and D. Weitz, Physical
  Review Letters {\bf 97},  215502  (2006).

\bibitem{Schm+94}
V. Schmitt, F. Lequeux, A. Pousse, and D. Roux, Langmuir {\bf 10},  955
  (1994).

\bibitem{radulescu99b}
O. Radulescu and P.~D. Olmsted, Rheol. Acta {\bf 38},  606  (1999).

\bibitem{SYC96}
N.~A. Spenley, X.~F. Yuan, and M.~E. Cates, J. Phys. II (France) {\bf 6},  551
  (1996).

\bibitem{LOB00}
C.~Y.~D. Lu, P.~D. Olmsted, and R.~C. Ball, Phys. Rev. Lett. {\bf 84},  642
  (2000).

\bibitem{doiedwards}
M. Doi and S.~F. Edwards, {\em The Theory of Polymer Dynamics} (Clarendon,
  Oxford, 1989).

\bibitem{SDO88}
T. Shimada, M. Doi, and K. Okano, J. the Physical Society of Japan {\bf 57},
  2432  (1988).

\bibitem{Cate96}
M.~E. Cates, J. Phys. Cond. Matt. {\bf 8},  9167  (1996).

\bibitem{Lopez-Gonzalez.Holmes.ea04}
M.~R. Lopez-Gonzalez, W.~M. Holmes, P.~T. Callaghan, and P.~J. Photinos, Phys.
  Rev. Lett. {\bf 93},  268302  (2004).

\bibitem{LerDecBer00}
S. Lerouge, J.~P. Decruppe, and J.~F. Berret, Langmuir {\bf 16},  6464  (2000).

\bibitem{Becu.Manneville.ea04}
L. Becu, S. Manneville, and A. Colin, Phys. Rev. Lett. {\bf 93},  018301
  (2004).

\bibitem{CatHeaAjd02}
M.~E. Cates, D.~A. Head, and A. Ajdari, Phys.\ Rev.\ E {\bf 66},  025202
  (2002).

\bibitem{Aradian.Cates05}
A. Aradian and M.~E. Cates, Europhys. Lett. {\bf 70},  397  (2005).

\bibitem{fielding04}
S.~M. Fielding and P.~D. Olmsted, Phys. Rev. Lett. {\bf 92},  084502  (2004).

\bibitem{Ganapathy.Sood06}
R. Ganapathy and A. Sood, Phys. Rev. Lett. {\bf 96},  108301  (2006).

\bibitem{Ganapathy.Sood06*1}
R. Ganapathy and A. Sood, Langmuir {\bf 22},  11016  (2006).

\bibitem{chaos2000}
R. Bandyopadhyay, G. Basappa, and A.~K. Sood, Phys. Rev. Lett. {\bf 84},  2022
  (2000).

\bibitem{SMF2005}
S.~M. Fielding, Phys. Rev. Lett. {\bf 95},  134501  (2005).

\bibitem{fielding06}
S.~M. Fielding and P.~D. Olmsted, Phys. Rev. Lett. {\bf 96},  104502  (2006).

\bibitem{WilsonFielding2006}
H. Wilson and S. Fielding, J. Non-Newt. Fl. Mech. {\bf 138},  181  (2006).

\bibitem{olmsted99a}
P.~D. Olmsted, O. Radulescu, and C.-Y.~D. Lu, J. Rheology {\bf 44},  257
  (2000).

\bibitem{numC}
W. Press, S.~A. Teukolsky, W.~T. Vetterling, and B.~P. Flannery, {\em Numerical
  Recipes in C}, 2nd edition ed. (Cambridge University Press, Cambridge, 1992).

\bibitem{grindrod}
P. Grindrod, {\em The Theory and Applications of Reaction-Diffusion Equations
  {\textit Patterns and Waves}} (Clarendon Press, Oxford, 1996).
\bibitem{yih} C.-S. Yih, J. Fluid Mech. (1967), {\bf 29}, 539-544 (1967).

\bibitem{lapack} Anderson, E. and Bai, Z. and Bischof, C. and
                Blackford, S. and Demmel, J. and Dongarra, J. and
                Du Croz, J. and Greenbaum, A. and Hammarling, S. and
                McKenney, A. and Sorensen, D. \em{{LAPACK} Users' Guide},
                (Society for Industrial and Applied Mathematics, Philadelphia, PA, 1999)

\end{thebibliography}
\end{document}